\documentclass[useAMS,usenatbib]{mn2e}
\usepackage{graphicx}
\usepackage{amsmath,amssymb}
\usepackage[colorlinks,citecolor=blue,urlcolor=blue,linkcolor=blue]{hyperref}

\def\aj{AJ}%
\def\araa{ARA\&A}%
\def\apj{ApJ}%
\def\apjl{ApJ}%
\def\apjs{ApJS}%
\def\apss{Ap\&SS}%
\def\aap{A\&A}%
\def\cjaa{Chinese J. Astron. Astrophys.}%
\def\mnras{MNRAS}%
\def\pasj{PASJ}%
\newcommand{\drj}{^{\circ}}
\newcommand{\nbody}{\textsc{nbody6}}
\newcommand{\notide}{\textsc{notide}}
\newcommand{\draco}{\textsc{draco}}
\newcommand{\msun}{M_{\odot}}
\newcommand{\mstar}{M_{\ast}}
\newcommand{\rvir}{R_{\rm vir}}

\title[The dynamical fate of binary star clusters]{The dynamical fate of binary star clusters in the Galactic tidal field}

\author[Priyatikanto, Kouwenhoven, Arifyanto, Wulandari \& Siregar]{
R. Priyatikanto,$^{1,2}$\thanks{E-mail: rho2m@hotmail.com}
M.B.N. Kouwenhoven,$^{3}$\thanks{E-mail: kouwenhoven@pku.edu.cn}
M.I. Arifyanto,$^{4}$
\newauthor{
H.R.T. Wulandari$^{4}$
and S. Siregar$^{4}$}
\\
$^{1}$Department of Astronomy, Faculty of Mathematics and Natural Sciences, Institut Teknologi Bandung, Bandung, Indonesia\\
$^{2}$Space Science Centre, National Institute of Aeronautics and Space, Bandung, Indonesia\\
$^{3}$Kavli Institute of Astronomy and Astrophysics, Peking University, Yiheyuan Lu 5, Haidian District, Beijing 100871, P.R. China\\
$^{4}$Astronomy Research Division, Faculty of Mathematics and Natural Sciences, Institut Teknologi Bandung, Bandung, Indonesia
}

\begin{document}

\date{}


\maketitle

\label{firstpage}


\begin{abstract}
Fragmentation and fission of giant molecular clouds occasionally results in a pair of gravitationally bound star clusters that orbit their mutual centre of mass for some time, under the influence of internal and external perturbations.
We investigate the evolution of binary star clusters with different orbital configurations, with a particular focus on the Galactic tidal field.  We carry out $N$-body simulations of evolving binary star clusters and compare our results with estimates from our semi-analytic model. The latter accounts for mass loss due to stellar evolution and two-body relaxation, and for evolution due to external tides. Using the semi-analytic model we predict the long-term evolution for a wide range of initial conditions. It accurately describes the global evolution of such systems, until the moment when a cluster merger is imminent. $N$-body simulations are used to test our semi-analytic model and also to study additional features of evolving binary clusters, such as the kinematics of stars, global cluster rotation, evaporation rates, and the cluster merger process. We find that the initial orientation of a binary star cluster with respect to the Galactic field, and also the initial orbital phase, are crucial for its fate. 
Depending on these properties, the binaries may experience orbital reversal, spiral-in, or vertical oscillation about the Galactic plane before they actually merge at $t\approx100$~Myr, and produce rotating star clusters with slightly higher evaporation rates. The merger process of a binary cluster induces an outburst that ejects $\sim10\%$ of the stellar members into the Galactic field.
\end{abstract}



\section{Introduction}

Stars emerge from giant molecular clouds through contraction and fragmentation. Most stars are formed in crowded stellar environments \citep{lada2003,kroupa2005,kruijssen2012} which belong to  star formation complexes ranging in size from sub-parsec to multi-kiloparsec scales \citep{elmegreen1996}. As shown by, e.g., \cite{allison2009b,allison2009a}, these systems are often highly substructured and can rapidly evolve into star clusters with smooth density profiles \citep[however, see also][]{banerjee2015a}. Many star clusters live short \citep[$t \lesssim 30$ Myr; see, e.g.,][]{degrijs2007} before being disrupted due to the high rate of destruction resulting from both internal and external processes \citep{marcos2008}, leaving moving groups and stellar streams at larger scales or interacting binary star clusters at smaller scales \citep{marcos2009}.

The existence of binary star clusters in nearby galaxies became of topic of interest when \cite{bhatia1988} published a list of 69 star cluster pairs in the Large Magellanic Cloud (LMC) with projected separations of less than 18~pc. Their statistical analysis demonstrated that over half of these cluster pairs are physically interacting, rather than just chance projections. The catalog was appended by \cite{pietrzynski1999b} who identified 73 double clusters, 18 triples, and 5 quadruples from the Optical Gravitational Lensing Experiment (OGLE) mission. \cite{dieball2002} suspected 473 clusters in LMC as the component of double or multiple systems. In the Small Magellanic Cloud (SMC), \cite{pietrzynski1999a} listed 23 doubles and 4 triples. Most of the cluster pairs have projected separations less than 20~pc, have relatively young ages, and their components have small age differences, suggesting that they are formed together as primordial binary star clusters \citep{dieball2002}.

In contrast to the LMC, binary star clusters appear to be less common in the Milky Way galaxy. Among hundreds of star clusters in the Solar neighbourhood, \cite{subramaniam1995} identified 16 cluster pairs with spatial separation less than 20 pc. Based on the updated WEBDA catalogue \citep{mermilliod2003} and the New Catalog of Optically Visible Open Clusters and Candidates \citep[NCOVOCC; see][]{dias2002}, \cite{marcos2009} proposed a list of 34 cluster pairs with 11 pairs having minor age differences. Several among these pairs, such as NCG 1981-NGC\,1976 and NGC\,3293-NGC\,3324, are considered as weakly interacting systems, while NGC\,6613-NGC\,6618 is a semi-detached system with an estimated combined tidal radius of the primary component that exceeds their physical separation. The lower prevalence of binary star clusters in the Milky Way (with respect to the younger LMC) is most likely a result of the relatively short lifetime of binary star clusters. In addition, the difference may also be partially attributed to a denser environment or a stronger tidal field exerted by the Galaxy or possibly even a lower frequency of binary cluster formation.

Binary star clusters are short-lived. Clusters in tight binaries experience mutual tidal interactions that may disrupt the smaller of the two clusters, or may result in a star cluster merger. Binary clusters in wide orbits suffer from interaction with the galactic tidal field that often breaks up the gravitational bond and separates the binary components \citep{marcos2010}. For binary star clusters in the LMC, \cite{bhatia1990} predicted that the lifetimes of the binaries are in the order of $10-100$ Myr, by considering merger lifetimes of binary clusters with different separations, and the disruption rate resulting from the galactic tidal field and passing giant molecular clouds. However, binary clusters can be more stable, i.e., having longer lifetimes, when they are in a retrograde orbit with respect to their revolution around the host galaxy \citep{innanen1972}. Although \cite{innanen1972} considered cases of binary satellite galaxies in an eccentric Galactic orbit, they demonstrated that the emerging Coriolis force may influence the orbits of the binary satellites and determines the stability of the systems.

The effect of the Galactic tidal field on the orbital evolution of binary star clusters has been studied by \cite{marcos2010}. These authors mention three possible fates of binary clusters: mergers, separations, and disruptions (shredded secondaries). Prior to the merger, the binary clusters follow complicated trajectories, different from the typical inspirals that are seen in simulation of mergers of isolated (i.e., without an external tidal field) binary  star clusters \citep{sugimoto1989,makino1991}.

Previous studies only considered one class of galactic orbits: those that are co-planar (and almost exclusively prograde) with respect to the Galactic plane. Given the crucial role of the Galactic tidal field on binary star cluster evolution, it is of interest to explore other orbital configurations that may affect the lifetimes of these systems. Moreover, merging binary clusters yield single star clusters with significant rotation \citep{makino1991}, and different initial orbital configurations may result in different types of rotating star clusters. The subsequent dynamical evolution of these rotating merger remnants is an important topic by itself \citep[e.g.,][]{einsel1999, kim2002, kim2004, ernst2007}, but additionally depends on the clusters' orientations with respect to their Galactic orbits, which can vary considerably, depending on the orientation of the binary orbit prior to the merger.

The formation process of star clusters is highly complex process. Primordial binary star clusters considered here are assumed to be in virial equilibrium and devoid of residual gas from their formation. Expulsion of the gas within roughly a million years dilutes the gravitational potential and determines the survivability of the clusters \citep[see, e.g.,][and references therein]{kroupa2001, pfalzner2013a, pfalzner2013b, banerjee2014}. This early phase of evolution is typically followed by a re-virialisation process that takes place at a timescale ranging between roughly 1~Myr and 10~Myr, depending on the mass of the cluster \citep{banerjee2013, pfalzner2014, banerjee2015b, banerjee2015a}. These processes are not covered in the present study. The zero point for the time, as used in this paper, can therefore be considered as the time at the clusters are re-virialised. Finally, we assume that all star clusters modelled in this paper have no spin at $t=0$, which may be a reasonable assumption, as even if a cluster were initially rotating, then gas expulsion, violent relaxation, and re-virialisation are likely to have removed most net angular momentum during the earliest phase cluster evolution.
Although not covered in this paper, the processes of gas expulsion and early dynamical evolution remain exciting topics that require a follow-up investigation.

In this paper we study the dynamical evolution of binary star clusters and their subsequent merger products in the Galactic tidal field using $N$-body simulations, and we develop a semi-analytic model. We provide a theoretical background regarding the dynamical evolution of binary star clusters in \autoref{theory}. We subsequently discuss the semi-analytic model in \autoref{semianalytic} and the initial conditions for the $N$-body simulations in \autoref{methods}. The results and analysis are provided in \autoref{results}. We discuss our results in \autoref{discussion} and finally summarise our conclusions in \autoref{conclusion}.


\section{Theoretical background} \label{theory}

The evolution of binary star clusters is influenced by both internal and external processes. The internal factors to be considered are cluster mass loss and evaporation \citep[e.g.,][]{spz2007}, tidal synchronisation \citep[e.g.,][]{sugimoto1989} and dynamical friction \citep[e.g.,][]{binney2008}. In the case of a binary stellar system, the 'term dynamical friction' is used to describe the dissipation of the clusters' orbital energy into the clusters' internal energy. The external processes responsible for their evolution are the tidal field of parent galaxy, as well as perturbations by giant molecular clouds \citep[e.g.,][]{bhatia1990}. All these processes affect the evolution of the binary orbit through a complicated and interrelated way, primarily through the transfer of angular momentum and energy.

When we describe the dynamics of binary star clusters as a two-body problem with clusters of mass $M_1$ and $M_2$, then the angular momentum of such an orbit with semi-major axis $a$ is
\begin{equation}
L=\mu a^2\omega=\dfrac{q(M_1+M_2)}{(1+q)^2}a^2\omega,
\end{equation}
where $\mu=(M_1M_2)/(M_1+M_2)$ is the reduced mass of the system, $0<q=M_2/M_1\leq 1$ the mass ratio, and $\omega$ the orbital angular velocity. 
From this equation we can see that, when the total angular momentum is conserved, a decreasing total mass, for example resulting from stellar mass loss, yields an increasing semi-major axis. Decreasing orbital angular momentum (with constant mass), for example as a result of the tidal torque, synchronisation or dynamical friction, results in orbital shrinkage (i.e., binary hardening), and in an increase of the angular velocity, as $\omega\propto a^{-3/2} \propto L^{-3}$.
Dynamical friction, the effect of which is strongest during a close encounter, converts some of the orbital kinetic energy into internal energy of the clusters, which decreases the angular speed. Angular momentum conservation consequently results in an expansion the orbit (under the assumption that a negligible amount of the angular momentum is stored in the internal rotation of the two star clusters). These basic processes, all of which affect the evolution of binary star clusters, are briefly reviewed below. 


\subsection{Cluster mass loss}

Star clusters lose their mass mainly through stellar evolution, ejections following two-body encounters, evaporation, and tidal disruption \citep{lamers2005}. Mass loss resulting from stellar evolution during the early phase of star cluster evolution has an important influence on the orbital evolution of binary star clusters. Depending on its initial mass function (IMF), a star cluster may lose about 20\% of its mass during the first tens of millions of years \citep[see, e.g.,][and references therein]{chernoff1990,kouwenhoven2014}. As demonstrated by \cite{spz2007} for binary star clusters without external tidal fields, mass loss tends to widen the orbital separation and also alter the orbital eccentricity. When mass loss occurs near the pericentre, it increases the orbital eccentricity, while mass loss near the apocentre tends to circularise the orbit.
Stellar evolution becomes less important as clusters become older. At this stage, cluster mass loss due to evaporation dominates the mass loss rate. Due to two-body interactions, some stars gain more energy and may escape into the galactic field as their energy exceeds the binding energy of the system. As the clusters continue to lose mass, the tidal field gradually reduces the energy limit for escaping stars.

\begin{figure*}
\centering
\includegraphics[width=0.8\textwidth]{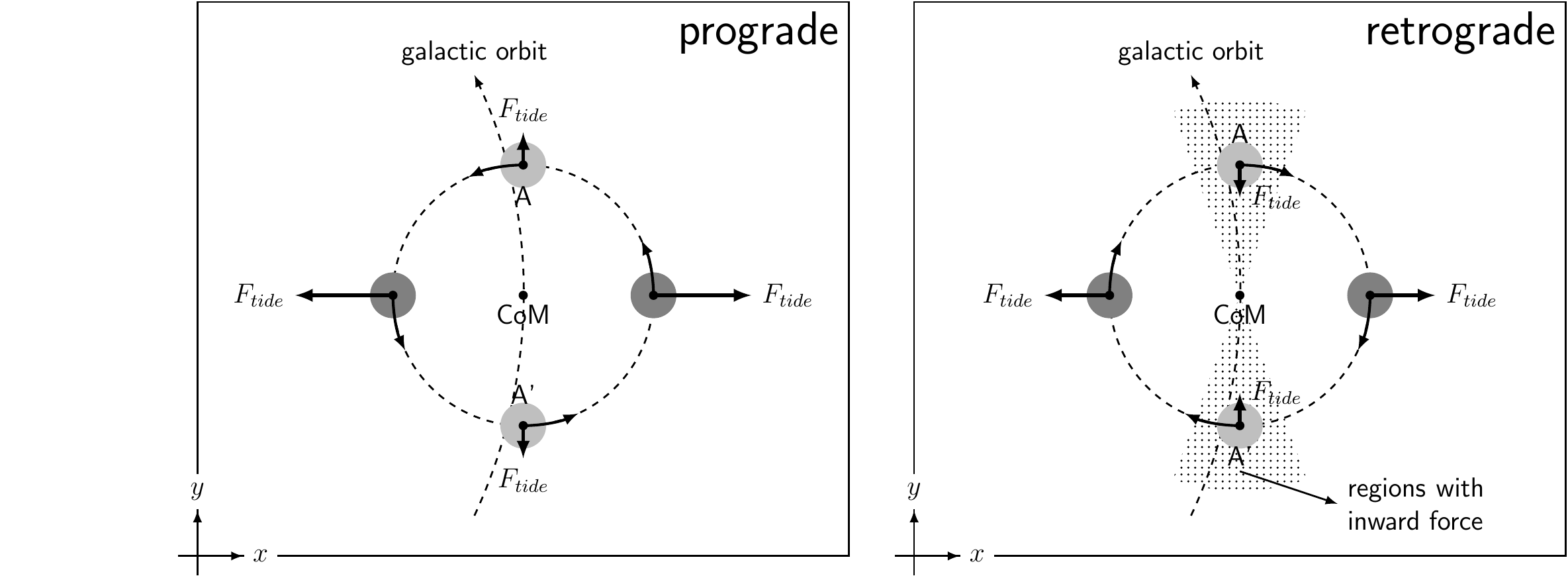}
\caption{The two components of a binary star cluster in the Milky Way experience a different net force, depending on their positions and velocities relative to the centre of the galaxy (see \autoref{eq:torque}). Along circular orbits at different radii, the net vectors of the tidal force vary as a function of position and velocity. For retrograde orbits (right-hand panel) there are regions where the net force acts as in inward force (shaded regions) that shrinks the orbit. For example, when the two clusters are at points A and A', respectively, there is no difference in gravitational potential, but the Coriolis force points towards the binary clusters' centre of mass. To quantify the effect of these forces, we investigate the evolution of binary star clusters with different initial configurations (see \autoref{initial}).  }
\label{torquex}
\end{figure*}

Stars escaping as a result of evaporation or ejection typically carry away angular momentum, and this generally results in a shrinkage of the binary cluster orbit. \cite{sugimoto1989} quantify the angular momentum loss rate by escaping stars of mass $\Delta m$ as
\begin{equation}
\frac{ \Delta L_{\text{esc}} }{\Delta m} =-la^2\omega \ ,
\label{angmo}
\end{equation}
where $l$ is a dimensionless constant representing the specific angular momentum taken away by escaping stars. The latter is related to the directions into which the stars escape, which is generally non-isotropic due to the presence of an external Galactic tidal field. The value of $l$ was the subject of analysis by some authors \citep[e.g.,][]{nariai1975} for the case of binary stars, in order to describe particles escaping through one of the Lagrange points ($L_2$). The typical value of $l$ is about 1.7 for escaping gas from a binary star, while for the case of stellar system in binary orbit, \cite{sugimoto1989} find $l=1.2$.
Angular momentum loss by escaping stars plays a crucial role during the merger process of binary star clusters and the subsequent merger products. Escaping stars take away angular momentum and hence reduce the orbital separation to the extent where tidal synchronisation and tidal friction become important. Without considering this effect, the semi-analytic models of \cite{spz2007} tend to result in separation of the two binary cluster components, rather than in a merger.


\subsection{Tidal synchronisation}

Similar to the case of inspiraling binary stars, the mutual tidal force plays an important role in the evolution of binary star clusters. As discussed in \cite{sugimoto1989}, star clusters in a binary orbit with initially zero (internal) rotation obtain spin angular momentum from the orbit through tidal interaction with the other companion. Angular momentum transfer is required in order to synchronise the rotation and revolution. Consequently, the orbital angular momentum decreases and the orbital separation shrinks. This synchronisation process can result in an instability, as the rate of angular momentum transfer increases. This occurs when the orbital separation reaches a critical separation $r_{\rm crit}$. For the equal-mass case,  $r_{\rm crit}\approx\sqrt{12}\,\mathcal{R}_g$, where $\mathcal{R}_g^2=\sum m_ir_i^2/\sum m_i$ denotes the gyration radius of the star cluster with respect to its rotation axis.


\subsection{Dynamical friction}

Close encounters between two stellar systems are accompanied by dynamical friction, where part of the orbital kinetic energy is converted to internal kinetic energy. As a result, the binary cluster's orbital velocity, the angular momentum, and the orbital separation are reduced, while the radii of the two clusters grow as a response to the increasing internal energy.
Dynamical friction is the key process that results in star cluster mergers, although a quantitative comparison to the synchronisation instability has not yet been explored. Many authors have discussed this process as the responsible factor for orbital decay of satellite galaxies \citep[e.g.][]{gardiner1994}. For binary star cluster mergers, \cite{bhatia1990} argue that the merging timescale is in agreement with the prediction of tidal friction, particularly under the impulse approximation \citep[e.g.,][]{alladin1978, spitzer1987, binney2008}. The only major discrepancy between the two approaches occurs at small cluster separations, where the impulse approximation greatly underestimates the merger timescale.
In the impulse approximation, the relative change of the internal energy of a stellar system, primarily the kinetic energy, depends on the mass ($M_p$), the impact parameter ($b$), the velocity ($V$) of the perturbing body, and also on the cross-sectional area of perturbed body ($\propto R^2$). \cite{binney2008} suggest the proportionality
\begin{equation}
\dfrac{\Delta E}{|E|}\propto \left(\dfrac{M_p R}{b^2 V}\right)^2 \ 
\end{equation}
for the relative change in the energy.


\subsection{The Galactic tidal field}

The Galactic tidal field plays an important role in determining the fate of star clusters in the Solar neighbourhood \citep{lamers2006}. Provided that a star cluster survives the infant mortality phase, and is not harassed by giant molecular clouds, the external tidal field is the dominant contributing factor to the disruption of star clusters. The same applies to star clusters that are part of a binary cluster, and to the binary star cluster systems themselves. A static tidal field prunes stars in the cluster halo and sets a limit to the cluster's size \citep{king1966}. As mentioned above, the tidal field tends to separate clusters from their companions. However, under some conditions, the net effect of the tidal force acts as inward force that supports the merging process.

In the case of a star cluster with a circular orbit around the Galactic centre, the tidal field can be approximated using Hill's approximation, or the epicyclic approximation \citep{binney2008}. In this approximation, the star clusters are assumed to be much smaller than the distance to Galactic centre. In this scheme, it is convenient to express the position and velocity of stars in the rotating coordinate system centred on the cluster's (or binary cluster's) centre of mass. In this coordinate system, the Galactic centre is located at $(-R_\odot,0,0)$, where $R_{\odot}=8.5$~kpc. The coordinate system rotates with an angular speed of $\Omega_{0}=2.7\times10^{-2}$ Myr$^{-1}$.
The Oort constants $A=14.4$ km\,s$^{-1}$\,kpc$^{-1}$ and $B=-12.0$ km\,s$^{-1}$\,kpc$^{-1}$ and the local stellar density of $\rho_G=0.11\,\msun$\,pc$^{-3}$ for the Solar neighbourhood can be used to parameterise the Galactic tidal field, and $\Omega_{0}=A-B$. Then, any star that is located at position $(x,y,z)$ and moves with velocity $(v_x,v_y,v_z)$ experiences an external acceleration $(a_x,a_y,a_z)$ of
\label{eq:torque}
\begin{align} 
a_x &= 2(A-B)v_y + 4A(A-B)x, \nonumber\\
a_y &=-2(A-B)v_x,
\label{tidal}\\
a_z &=-\left[4\pi G\rho_G+2\left(A^2-B^2\right)\right]z \ , \nonumber
\end{align}
where $G$ is the gravitational constant. The star cluster is truncated near the tidal radius, which can be approximated with the Jacobi radius,
\begin{equation}
r_J\equiv\left[\dfrac{GM}{4A(A-B)}\right]^{1/3}.
\label{rtide}
\end{equation}
Stars located at the Lagrange points, including $(\pm r_J,0,0)$, experience no net acceleration. 

Since the magnitude and the direction of the external force depend on both position and velocity, the tidal field acts differently on prograde and retrograde binary cluster orbits. 
As illustrated in \autoref{torquex}, the tidal field always exerts an outward net force on a binary cluster in a prograde orbit (i.e., the tide is always extensive). In a retrograde orbit, on the other hand, there is an orbital segment where the external tide exerts a net inward force towards the centre of mass of the cluster system (i.e., the tide can under certain conditions be compressive). The portion of this segment depends on the radius and velocity of the orbit. This extensive or compressive tide is in fact the result Coriolis force experienced by the two moving clusters. This force is the result of a combination of two vectors: the angular rotation around the host galaxy ($\vec{\Omega}$) and the cluster's velocity ($\vec{v}$). The Coriolis component becomes prominent when the binary cluster components are positioned at points A and A' (\autoref{torquex}), because at those locations the difference in the external gravitational potential experienced by the two clusters vanishes. In a prograde system, the Coriolis force is directed away from the centre of mass, resulting in a extensive tide. In a retrograde orbit, on the other hand, the Coriolis force becomes a compressive force.
Therefore, we may expect that binary clusters with retrograde orbits have a larger probability of merging, while systems with prograde orbits are more vulnerable to the disruptive effect of the tidal field.


\section{Semi-analytic model}   \label{semianalytic}

As a bridge between theory and more realistic (but more complicated) $N$-body models, we develop a semi-analytic model that includes the aforementioned processes. The main purpose of this model is to predict the fate (merger or separation) of binary star clusters with different initial conditions. In this model, a star cluster is represented by a spherical body of mass $M_0$ and radius $R_0$ that belongs to initially bound binary orbit with a certain semi-major axis and eccentricity. The initial and orbital phase and orbital orientation with respect to the Galaxy can be adjusted to arbitrary values. A technical explanation about the set-up of the initial conditions can be found in \autoref{appendix1}.

The total mass of each star cluster changes over time due to stellar evolution and evaporation. The analytical prescription of \cite{lamers2005} is invoked to model these processes. The mass loss fraction due to stellar evolution, $q_{\text{ev}}\equiv (\Delta M)_{\text{ev}}/M_0$, can be approximated with
\begin{equation}
\log{q_{\text{ev}}}=\left(\log{t}-a_{\text{ev}}\right)^{b_{\text{ev}}}+c_{\text{ev}} \qquad\text{for } t>12.5\text{ Myr} \ ,
\end{equation}
where $a_{\text{ev}}$, $b_{\text{ev}}$ and $c_{\text{ev}}$ are constants that depend on the IMF and the metallicity. Following the GALEV stellar population model \citep{anders2003}, we use a \cite{salpeter1955} IMF with $\alpha=-2.35$, a mass range of 0.15 to 50 $\msun$ and a Solar metallicity, so that $a_{\text{ev}}=7.00$, $b_{\text{ev}}=0.255$ and $c_{\text{ev}}=-1.805$. As the Salpeter IMF is more bottom-heavy than the more realistic canonical IMF, the initial total cluster masses (given an identical number of member stars) are about 30\% lower than that for realistic star clusters. On the other hand, stellar and dynamical evolution result in a mass loss rate smaller than that for a cluster with a canonical IMF \citep[see][for a discussion on this topic]{kroupa2013}. Finally, we assume that the radius of each star cluster is constant over time.

At early times, cluster mass loss is dominated by stellar evolution. After several tens of millions of years, evaporation due to two-body relaxation and tidal stripping become the dominant mechanisms for mass loss. As discussed in \cite{lamers2005}, this mass loss rate depends on a combination of the total cluster mass and the strength of the external tidal field. This can be approximated analytically as
\begin{equation}
\left(\dfrac{dM}{dt}\right)_{\text{evap}}=\dfrac{\msun}{t_0}\left(\dfrac{M}{\msun}\right)^{1-\gamma},
\label{disrupt}
\end{equation}
where $t_0\approx 18$~Myr is a constant parameterising the lifetime of a star cluster in a given external tidal field.
The parameter $\gamma\approx0.6$ is an empirical mass loss scaling factor which is roughly constant \citep{boutloukos2003}. 

Combining the contributions of stellar evolution and dynamical mass loss, the evolution of the total mass of a star cluster can be approximated as
\begin{equation}
\dfrac{M(t)}{M_0}\simeq\left[(1-q_{\text{ev}})^{\gamma}-\dfrac{\gamma t}{t_0}
\left(\dfrac{\msun}{M}\right)^{\gamma}\,\right]^{1/\gamma} \ .
\label{massloss}
\end{equation}

The equation of motion of the binary cluster is integrated using the Hermite predictor-corrector scheme \citep{makino1992}, which includes gravitational attraction and the external tidal field. The time step for integration is $10^{-3}P$, where $P$ is initial orbital period of the binary. Both accelerations and their first derivatives are calculated. For each time interval, the mass loss is calculated using \autoref{massloss}, while angular momentum loss due to escaping stars is calculated using \autoref{angmo}, assuming $\Delta m$ to be mass loss due to relaxation and tidal stripping (\autoref{disrupt}). Instantaneous loss of an amount $\Delta L_{\text{esc}}$ of angular momentum is translated into a decrease in the tangential velocity ($v_t$) according to 
\begin{equation}
\dfrac{\Delta v_t}{v_t} = \dfrac{\Delta L_{\text{esc}}}{L}  \quad\quad\quad\Delta L_{\text{esc}}<0 \quad ,
\end{equation}
such that the new (instantaneous) tangential velocity is
$v_t' = v_t \left( 1+\Delta L_{\text{esc}}/L \right)$. Followed by circularisation, this process results in the gradual hardening of the binary star cluster, and can in the absence of external forces result in a merger between the two star clusters. 

The combined contributions of the mass loss, the angular momentum loss and the tidal torque determines the dynamical fate of the binary star cluster (merger or separation). When the separation of the clusters becomes small enough, the mutual tidal interaction becomes stronger and accelerates the merging process. However, our semi-analytic model does not include tidal synchronisation and dynamical friction which have reciprocal effect with the cluster radius (although these processes are well-modelled in the $N$-body model; see \autoref{methods}). Thus, the validity of our semi-analytic model is limited by the condition when the tidal effects become significant, i.e., the moment when the clusters approach each other's Roche lobes. We estimate the size of the Roche lobe is using 
\begin{equation}
R_{L}\approx\dfrac{0.44\,q^{0.33}}{(1+q)^{0.2}}\,r \ .
\label{rochelobe}
\end{equation}
\citep{eggleton1983}. Here, $q=M_2/M_1$ is the mass ratio of the clusters and $r$ is their mutual separation. If the radius of any cluster exceeds its Roche lobe, then the simulation is terminated and the binary system is expected to merge in the near future. This prediction is plausible since high-speed encounters between two clusters (without a subsequent merger) are very unlikely in the case of primordial binary clusters. On the other hand, when the separation between the two clusters exceeds their (combined) tidal radius within the Galactic field, then the separation continues to increase and the binary cluster becomes unbound.
In this paper we primarily concentrate on equal-mass binary clusters, for which the Roche radius is $R_L \simeq 0.38r$, where $r$ is the  distance between the clusters (see \autoref{rochelobe}). In our semi-analytic models we make the assumption that the cluster radii remain constant over time ($R=1.5$~pc), so the simulation is terminated at the moment when the orbital separation is less than $r=3.9$~pc. Simulation are also terminated when the separation between two clusters exceeds the (combined) tidal radius of the binary cluster system in the Galactic field (\autoref{rtide}).


\section{$N$-body simulations}   \label{methods}

\begin{table}
\caption{Initial conditions for the $N$-body simulations. Columns 1, 2, and 3 list the inclination angle, the longitude of the ascending node, and the initial orbital phase for the binary star clusters. The other initial conditions for the binary star clusters (\draco$1-7$ and \notide) are listed in column~4. The \textsc{single} star cluster consists of $N=4096$ stars and has a total mass of $M=2064~\msun$, while all other initial conditions are identical to those of the binary star clusters. }
\label{initial}
\renewcommand{\arraystretch}{1.2}
\begin{tabular}{l r r r |l}
\hline
Model & \multicolumn{1}{c}{$i$} & \multicolumn{1}{c}{$\Omega$} & \multicolumn{1}{c}{$\theta$} & Other parameters\\
\hline
\draco1 & $0\drj$   & $0\drj$  & $0\drj$  & $N=2\times2048$ \\
\draco2 & $180\drj$ & $0\drj$  & $0\drj$  & $M=2\times1032\,\msun$ \\
\draco3 & $0\drj$   & $0\drj$  & $90\drj$ & $\rvir = 1$~pc \\
\draco4 & $180\drj$ & $0\drj$  & $90\drj$ & $0.15\leq m\leq 50\,\msun$ \\
\draco5 & $90\drj$  & $0\drj$  & $0\drj$  & Salpeter (1995) IMF \\
\draco6 & $90\drj$  & $0\drj$  & $90\drj$ & $a=10$~pc \\
\draco7 & $90\drj$  & $90\drj$ & $0\drj$  & $e=0$ \\
{\notide} & $0\drj$ & $0\drj$  & $0\drj$  & Solar circle Galactic tide\\
\textsc{single} & $-$ & $-$ & $-$ & \\
\hline
\end{tabular}
\end{table}

We conduct $N$-body simulations of binary clusters in the Solar neighbourhood using the {\nbody} package \citep{aarseth2003}. Stellar evolution is included following the prescriptions of \cite{hurley2000}. In our models we adopt a solar metallicity ($Z=0.02$).
Each model consists of two identical star clusters with a Plummer distribution profile \citep{plummer1911}, each containing $N=2048$ single stars. To remain consistent with the initial conditions of the semi-analytic models (\autoref{semianalytic}), stellar masses are drawn from the \cite{salpeter1955} IMF with slope $\alpha=-2.35$ in the mass range $0.15\,\msun \leq \mstar \leq 50\,\msun$, resulting in an average stellar mass of $\langle \mstar \rangle \simeq 0.504\,\msun$ and an average total mass of $2064\,\msun$ of the binary cluster \citep[note that these values are smaller than for a canonical IMF; see also \autoref{semianalytic} and the review of][]{kroupa2013}. The individual star clusters are initialised in virial equilibrium with an initial virial radius of $\rvir=1$~pc, which corresponds to a half-mass radius of $0.77$~pc \citep[e.g.,][]{heggiehut}. 
All clusters are initially non-rotating, and for simplicity we do not include primordial binary stars.
The binary clusters are initialised on gravitationally-bound circular orbits with a semi-major axis of $a=10$ pc, which corresponds to an orbital period of $P=65.7$~Myr. 
During the simulation, star clusters are immersed in a Galactic tidal field identical to that of the Solar neighbourhood (see also \autoref{theory}). This external tidal field is adopted for comparison with the semi-analytic model, but can be adjusted for arbitrary environments \citep[see, e.g.,][]{renaud2011}. Note that this binary orbital period corresponds to approximately a quarter of the revolution period around the Galaxy for a binary cluster on a solar orbit. 

We study the evolution of binary clusters with different orbital orientations, which are defined by the inclination $i$, the longitude of the ascending node $\Omega$ and the phase angle $\theta$. The angle $\Omega$ is measured from the Galactic centre along the Galactic plane, while the inclination is the angle between the normal vector of the binary orbit and the normal vector of the Galactic plane. The phase angle $\theta$ usually measured from the pericentre of binary orbit, but for convenience we define phase angle zeropoint at the ascending node (see \autoref{orbital}). The initial conditions of the binary orbits are summarised in \autoref{initial} and illustrated in \autoref{torquex}. Their orbital orientations are in the $x-y$ plane ({\draco1}-4), in the $x-z$ plane (\draco5,6) or in the $y-z$ plane (\draco7). In order to quantify the contribution of external perturbations, we run for comparison also binary star clusters without an external tidal field ({\notide}), as well as single clusters (\textsc{single}).

During the merging process it is important to carry out a careful membership analysis before any further analysis, in particular when dealing with the orbital evolution of the binary system. To this end, we use the hierarchical clustering algorithm \citep{silvermann1996}, which assigns membership according to the nearest neighbour density estimate of each star ($\rho_i$). Stars with identical roots belong to the same cluster.
Subsequently, the centre of density $\vec{r}_{\rm cod}$ for each cluster is determined following the procedure of \citep{casertano1985},
\begin{equation}
\vec{r}_{\rm cod}=\dfrac{\sum \vec{r}_i\rho_i}{\sum\rho_i} \ ,
\end{equation}
and is used as the definition of each cluster centre. For kinematic analysis, it is more appropriate to use centre of mass of the innermost fraction of the member stars. The centre of mass position and velocity for each cluster are calculated using
\begin{equation}
\vec{r}_{\text{com}}=\dfrac{\sum \vec{r}_im_i}{\sum m_i} \quad\text{and}\quad
\vec{v}_{\text{com}}=\dfrac{\sum \vec{v}_im_i}{\sum m_i} \ ,
\end{equation}
where we only use the stars within a distance $R_{\rm lim}=\max(R_{25},R_{L})$ from the centre of density, where $R_{25}$ is the radius containing 25\% cluster mass, and $R_{L}$ is the Roche lobe of the cluster within binary system (\autoref{rochelobe}). The choice of this limiting radius is necessary to avoid false orbital element determination in a tidally distorted cluster during a close encounter with another cluster.
Using the centre of mass and the centre of velocity of each cluster, we determine the orbital elements of the binary cluster using the approximation for the two-body problem \citep[see, e.g.][]{kroupa2008}. 
The typical tidal radius of the system of mass $2064\msun$ is about $r_J\approx 18$~pc (\autoref{rtide}). Stars that escape from the binary cluster system and reach a distance beyond two Jacobi radii are removed from the system, and their kinematic properties are recorded.


\section{Results and analysis}
\label{results}

\begin{figure*}
\centering
\begin{tabular}{ccc}
\multicolumn{3}{c}{\includegraphics[width=0.80\textwidth]{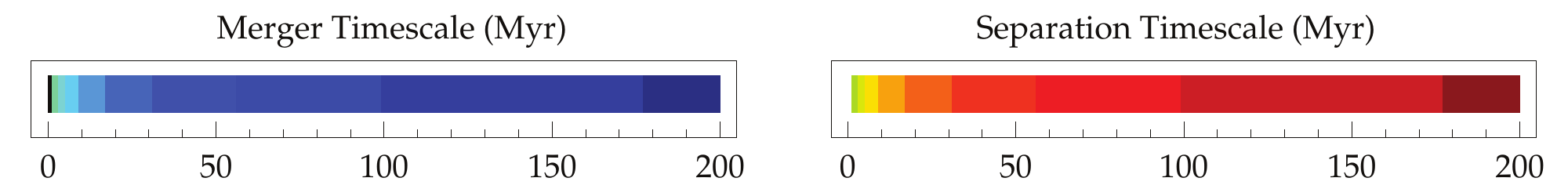}} \\
\includegraphics[width=0.32\textwidth]{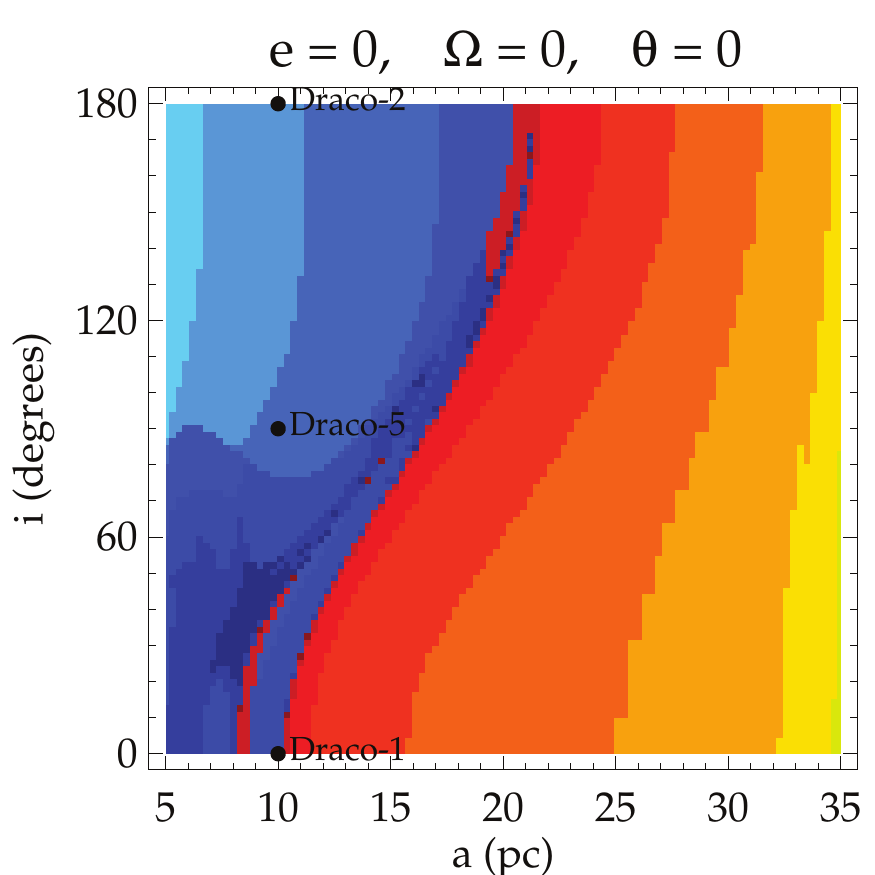} &
\includegraphics[width=0.32\textwidth]{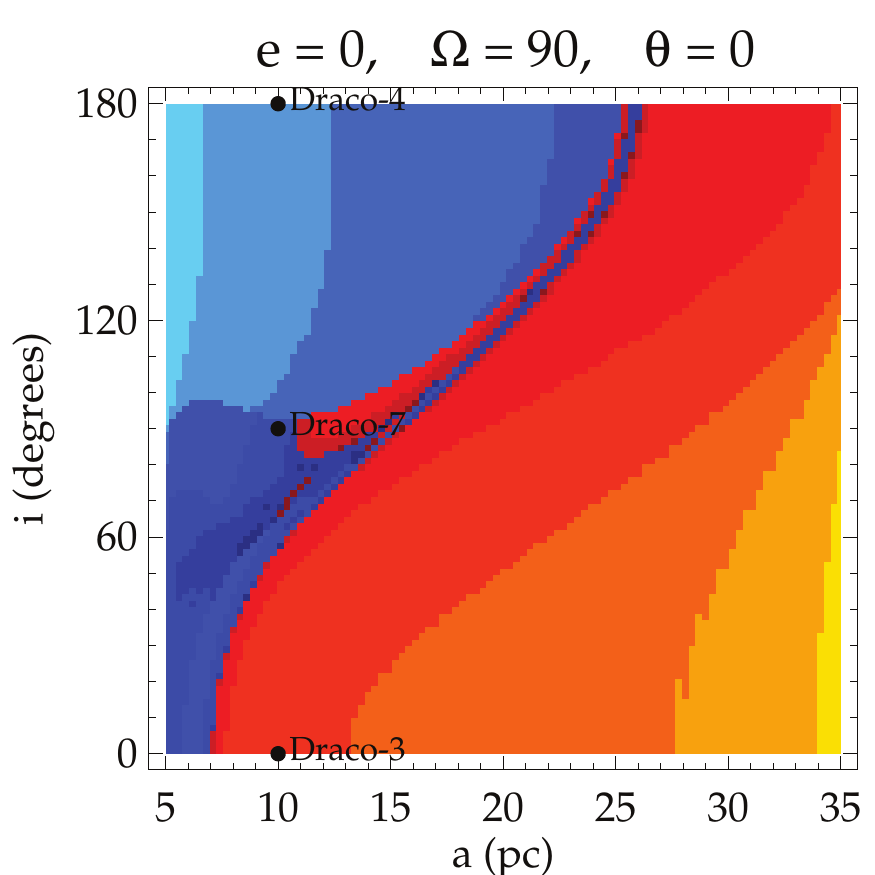} &
\includegraphics[width=0.32\textwidth]{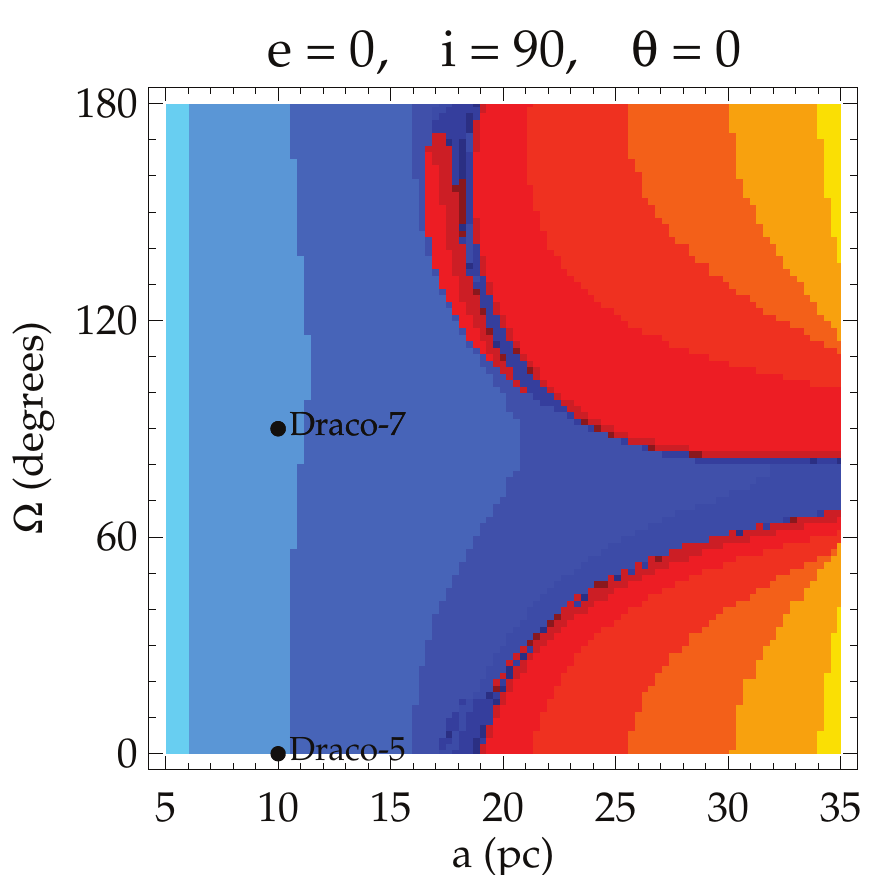} \\
\end{tabular}
\caption{Equal-mass binary clusters on circular orbits may experience merging or separation, depending on their initial orientation and orbital phase. The left-hand panel shows which binary clusters merge (blue) and which clusters separate (red), for different initial semi-major axes $a$ and inclinations $i$, but with a fixed initial longitude of ascending node $\Omega=0\drj$ (rotation about the $x$-axis). A merger is assumed to occur when the two clusters enter each other's Roche radius, and they are assumed to separate when their mutual separation exceeds the combined tidal radius of the system in the external galactic tidal field (see \autoref{semianalytic} for details).
The middle panel shows the results for $\Omega=90\drj$ (rotation about the $y$-axis) as a function of $a$ and $i$. The right-hand panel shows the results for $i=90\drj$ (rotation about the $z$-axis) as a function of $a$ and $\Omega$. The colours indicate the time at which a merger or separation occurs. Binary star clusters that lie near the boundary between the blue and red regions have relatively long lifetime. The initial condition for the $N$-body models are indicated with the black dots.}
\label{semi}
\end{figure*}


\subsection{Predictions from the semi-analytic model}

To characterise the influence of the tidal field on the evolution of a binary star cluster's orbit, we simulate models with different initial conditions. We restrict the simulations to binary clusters in circular orbits, but allow a range of initial values for the semi-major axis $a$, the orientation angles $\Omega$ and $i$, and the orbital phase $\theta$. Each binary cluster consists of two equal-mass components of $1032\,\msun$ and virial radii of $\rvir=1.7$~pc, which corresponds to a half-mass radius of 1.31~pc, and to a projected half-mass radius of 1~pc \citep[see, e.g.,][]{heggiehut}. These binary systems have a (combined) tidal radius of $r_J=17.9$~pc. We study binary clusters with initial semi-major axis ranging from $a=5$~pc, which is close to the merger limit, to $a=35$~pc, which is just below the combined tidal radius of the binary system.
\autoref{semi} shows the results of our semi-analytic models. The fate of a binary cluster (merger or separation) depends strongly on its initial orbital orientation with respect to the Galactic tidal field. 

Binary clusters with prograde orbits or low inclinations have a relatively large probability to be separated, as the tidal force always acts outwards (see \autoref{torquex}). In contrast, many of the systems with retrograde orbits, even with those with large initial separations, tend to shrink over time and ultimately merge. The left-hand and middle panels in \autoref{semi} show the results for binary clusters with identical physical and orbital properties, apart from the longitude of the ascending node $\Omega$. Although these results are qualitatively similar, the dynamical fate of many of these clusters (merger or separation) is ultimately determined by the initial value of $\Omega$ (see also the differences between models \draco1 and \draco3 in \autoref{position} and the discussion in \autoref{sec:nbodyev}), which is a direct result of the external tidal field. Another interesting prediction can be derived from the right-hand panel of \autoref{semi} for perpendicular orbits with different angles $\Omega$. There is a region near $\Omega\approx75\drj$ where the binary clusters, even when they have initial separations as wide as $a=35$~pc, are likely to merge.

Finally, we note that binary clusters that are initially located near the boundary between the merging binary clusters (blue region) and the separating binary clusters (red region) tend to survive longest. Binary clusters near this region experience a more or less balancing contribution from the processes that drive orbital shrinking and widening. The longest time span for our semi-analytic binary model is about 200~Myr. Although some of the assumptions, such as that of the time-independent radius, may no longer be appropriate at this age, our semi-analytic model gives us an order of magnitude estimate of the survival time of an equal-mass binary star cluster evolving within the tidal field of a galaxy. In general, the expected lifetime depends on many parameters, such as the masses of clusters and the mass ratio of the binary system, the initial separation, and the eccentricity. Among these possibilities, our model predicts the existence of relatively stable configurations, and direct $N$-body simulations can be used to explore these in further detail.


\subsection{Orbital evolutions of $N$-body models} \label{sec:nbodyev}

\begin{table}
\begin{minipage}{0.47\textwidth}
\centering
\caption{Merger time of various models (column~2) and the angular momentum components (in units of $10^3\msun$\,pc$^2$\,Myr$^{-1}$) of the merger remnants (columns~3, 4, and~5), measured $\sim10$~Myr after the merger has taken place. The uncertainty of the merger time is roughly $1$~Myr.}
\label{merger}
\renewcommand{\arraystretch}{1.2}
\begin{tabular}{ccrrr}
\hline
Model & Merger time & \multicolumn{3}{c}{Angular momentum}\\
& \multicolumn{1}{c}{(Myr)} & $L_x\ $ & $L_y\ $ & $L_z\ $\\
\hline
\draco1 & $92.2$ & $ 0.00$ & $-0.06$ & $-1.47$\\
\draco2 & $30.3$ & $ 0.25$ & $0.12$ & $-5.35$ \\
\draco3 & $\infty$ & $-$ & $-$ & $-$ \\
\draco4 & $49.5$ & $-0.12$ & $0.00$ & $-5.35$ \\
\draco5 & $44.5$ & $-2.27$ & $1.29$ & $-0.61$ \\
\draco6 & $62.8$ & $ 2.21$ & $-2.52$ & $-2.83$ \\
\draco7 & $65.9$ & $-1.60$ & $0.61$ & $-0.86$ \\
{\notide} & $131.8$ & $-0.06$ & $-0.06$ & $3.99$ \\
\hline
\end{tabular}
\end{minipage}
\end{table}

\begin{figure*}
\centering
\includegraphics[width=0.245\textwidth]{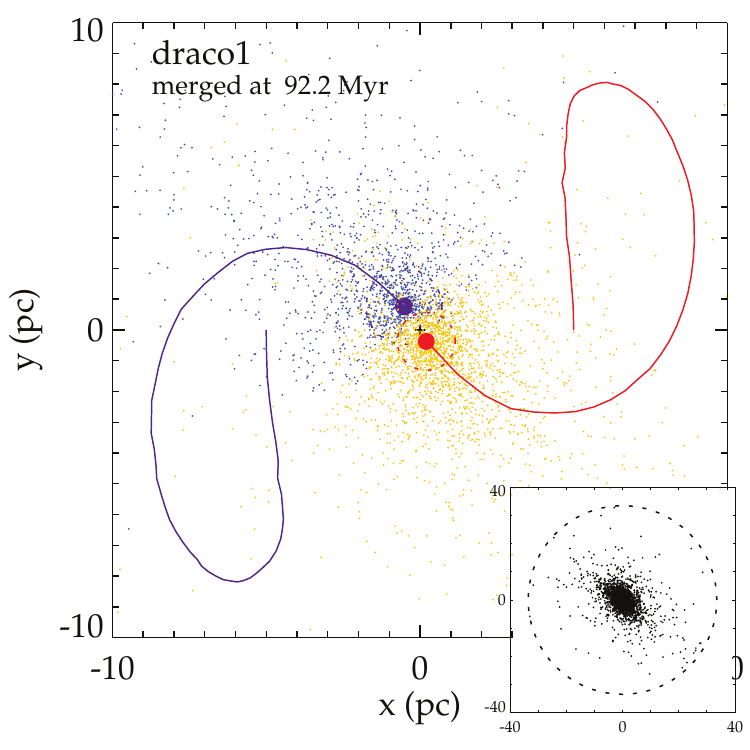}
\includegraphics[width=0.245\textwidth]{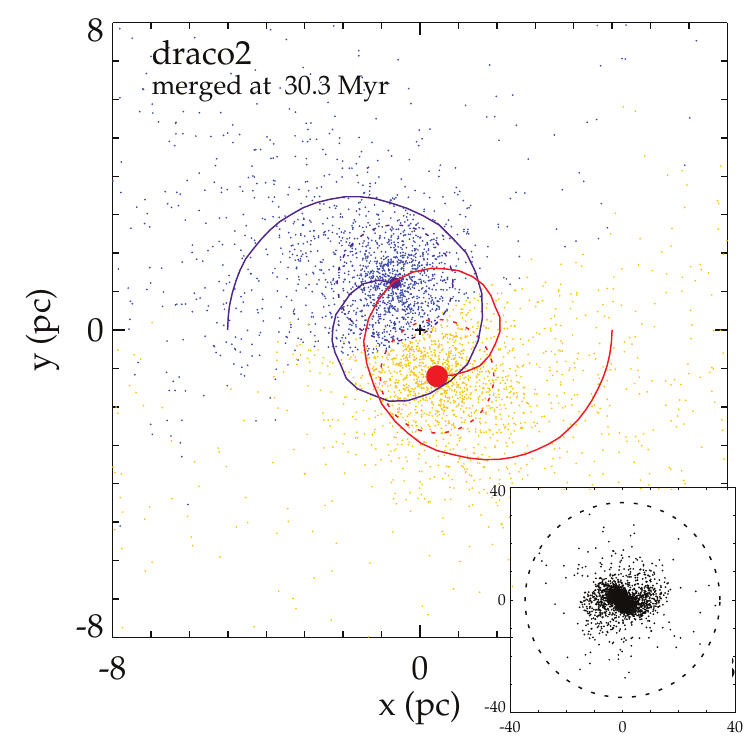}
\includegraphics[width=0.245\textwidth]{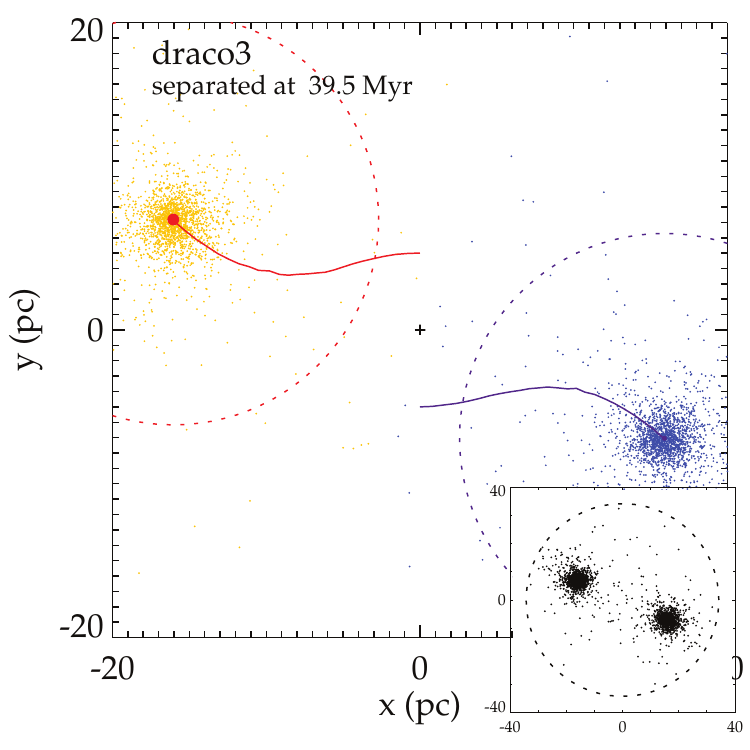}
\includegraphics[width=0.245\textwidth]{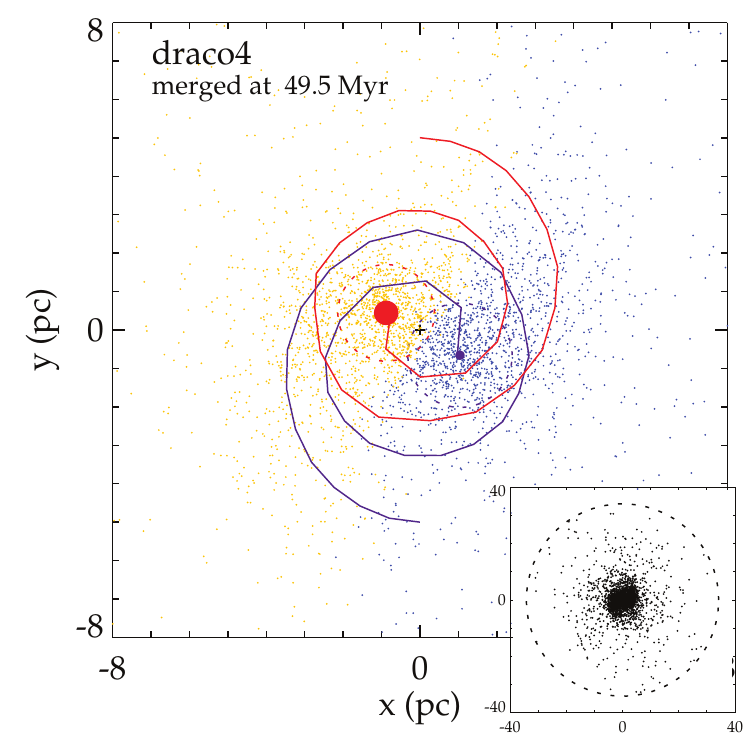}\\
\includegraphics[width=0.245\textwidth]{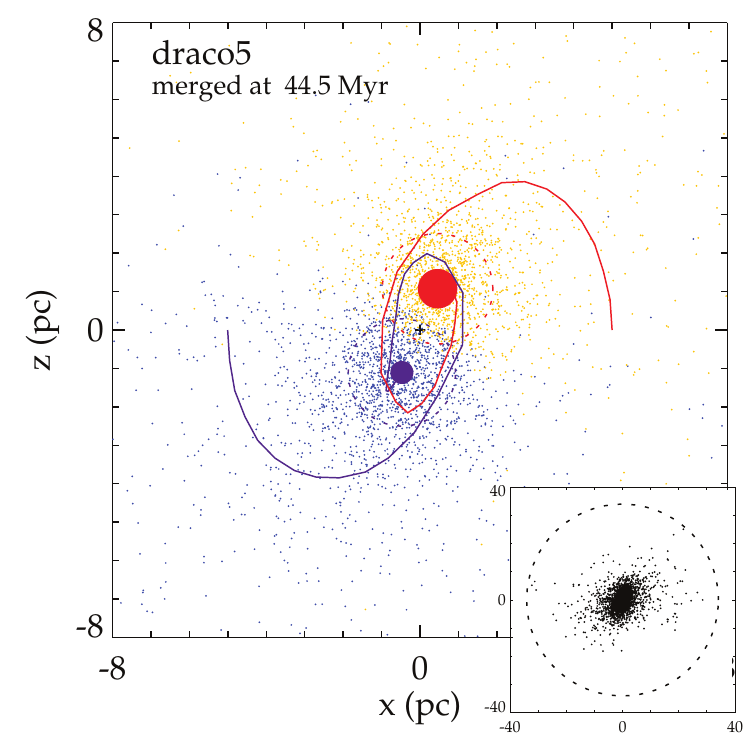}
\includegraphics[width=0.245\textwidth]{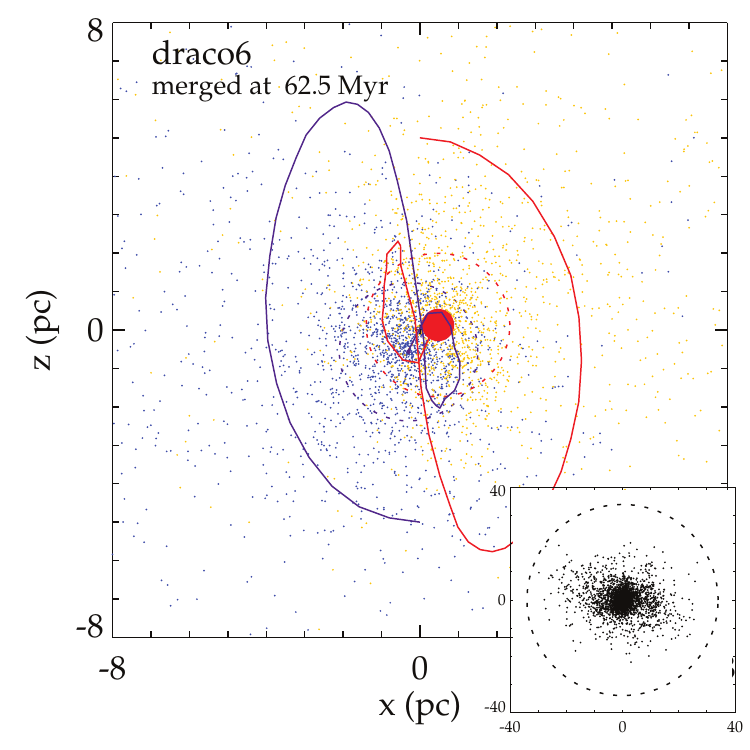}
\includegraphics[width=0.245\textwidth]{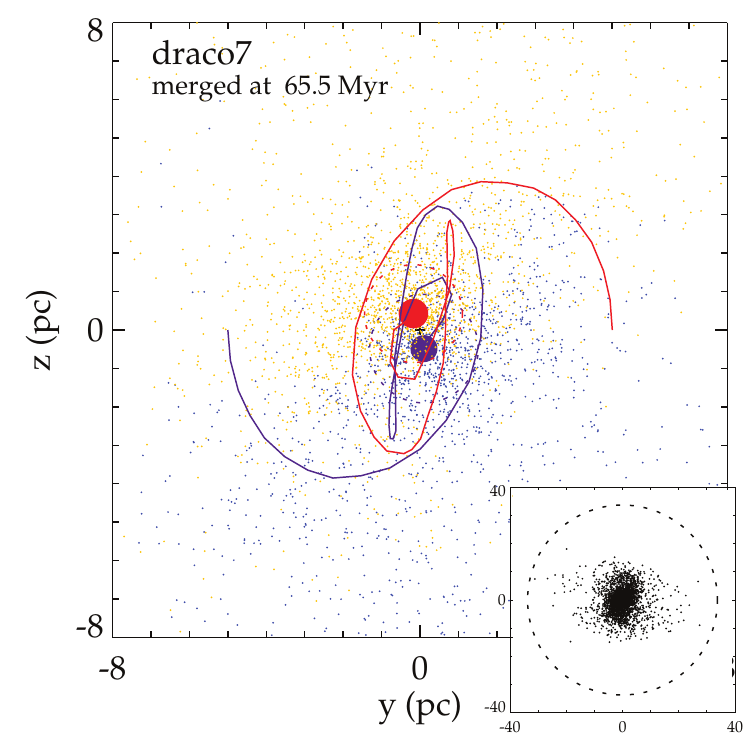}
\includegraphics[width=0.245\textwidth]{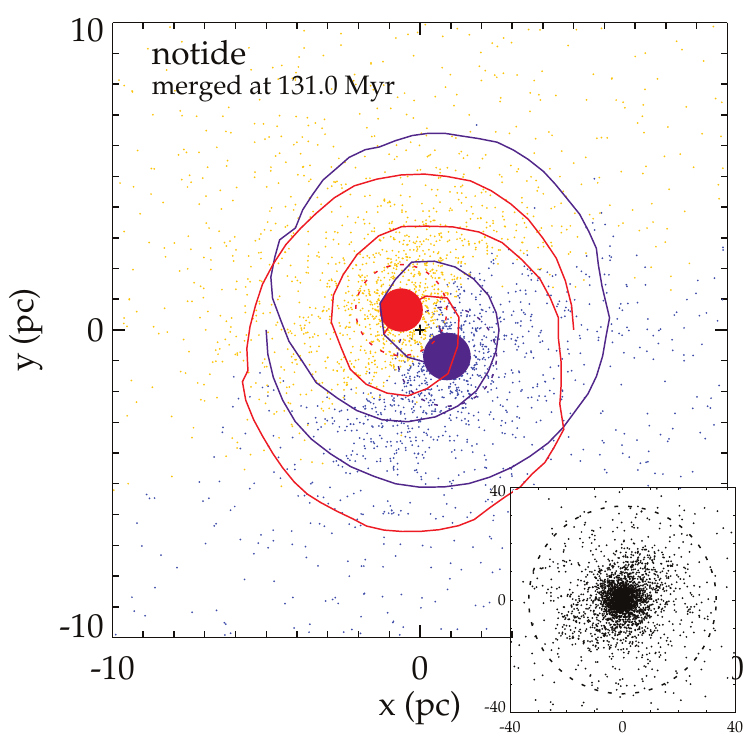}
\caption{Positions of the stars just before the binary star clusters merge (except for the \draco3 system, in which the two clusters separate). Each system is plotted in the orbital plane ($x-y$, $x-z$, or $y-z$) of their initial conditions. The blue and yellow dots indicate stellar membership to the two star clusters. The insets depict a wider field of view, where dashed circles mark the cut-off radius $r=2r_J\simeq 36$~pc. The blue and red curves mark the trajectories of the centres of mass (blue and red filled circles, respectively) of the two clusters since the start of the simulations.}
\label{position}
\end{figure*}

\begin{figure*}
\centering
\includegraphics[width=0.245\textwidth]{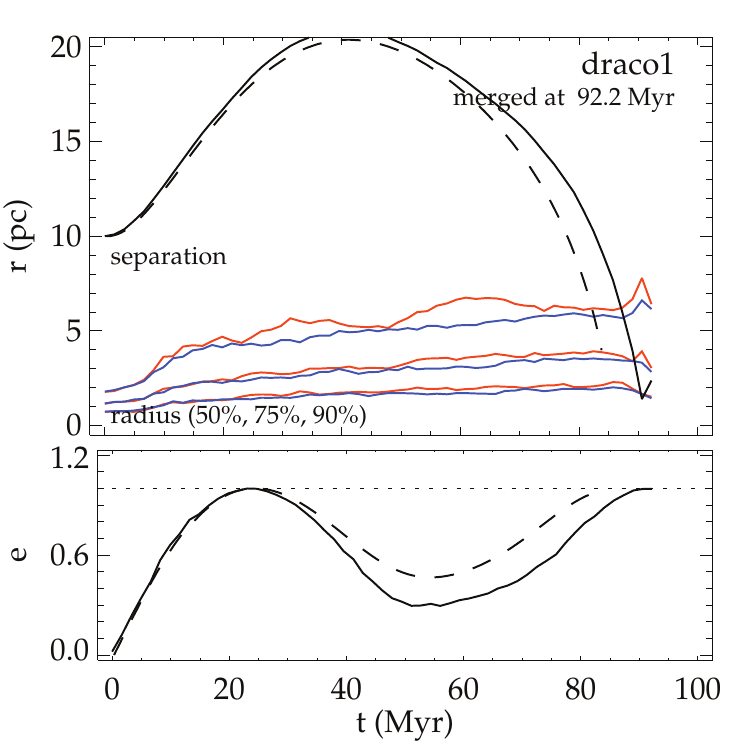}
\includegraphics[width=0.245\textwidth]{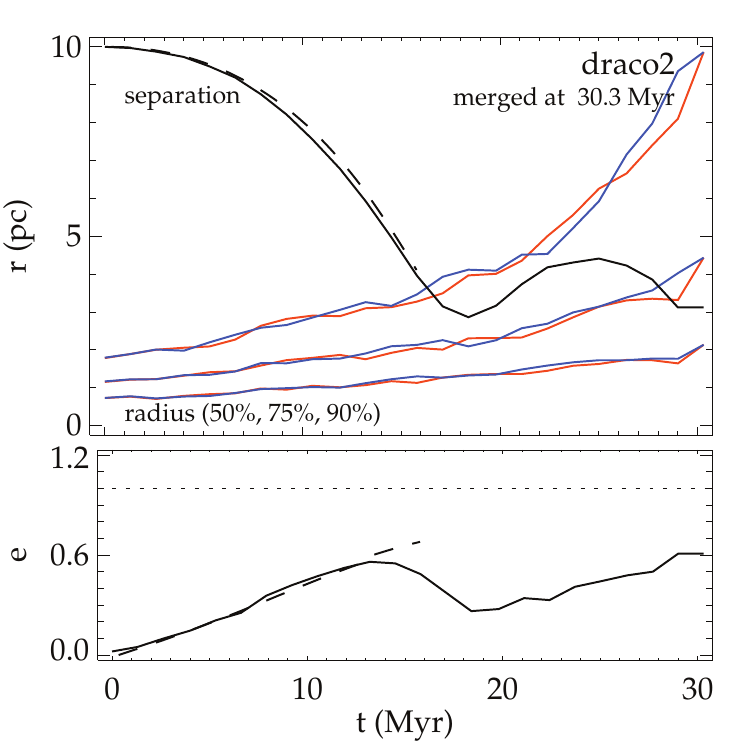}
\includegraphics[width=0.245\textwidth]{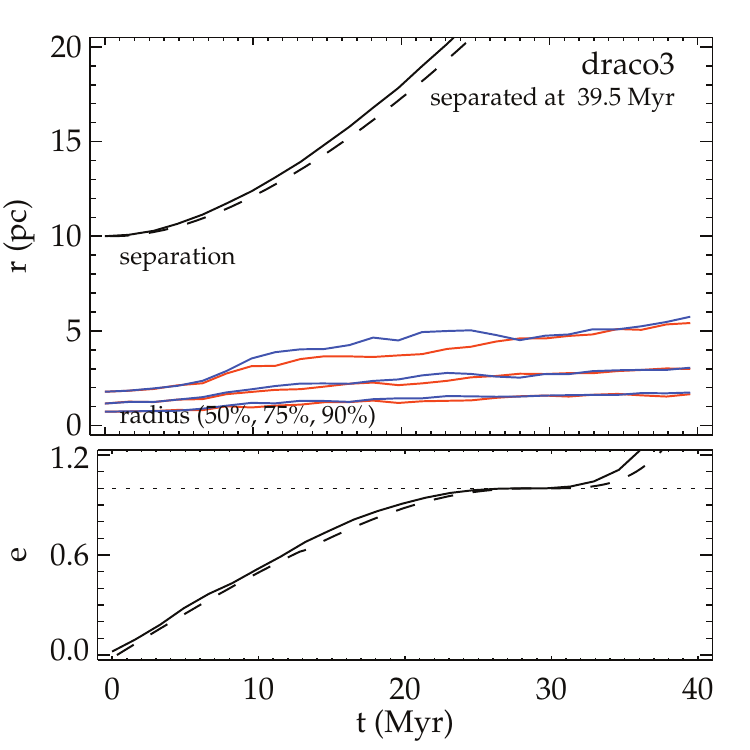}
\includegraphics[width=0.245\textwidth]{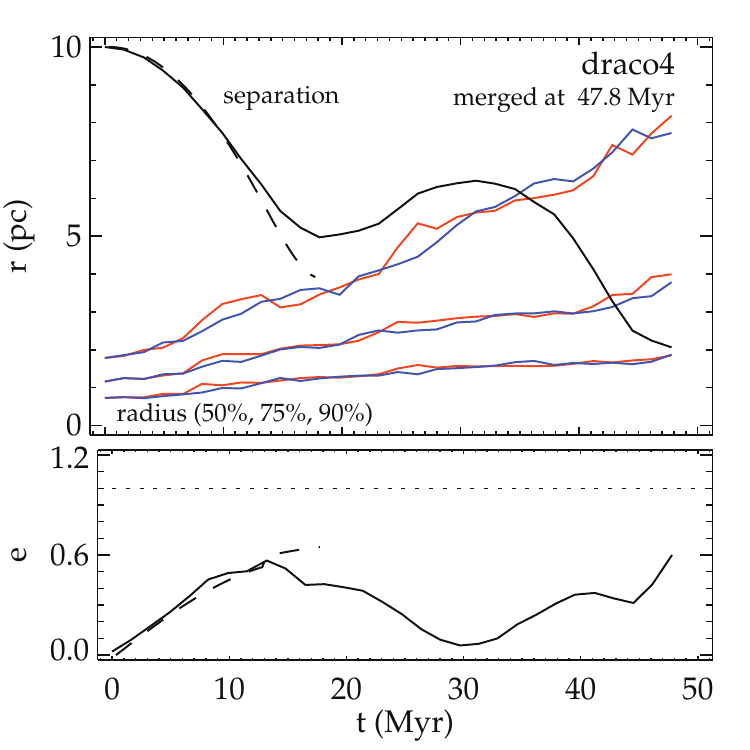}\\
\includegraphics[width=0.245\textwidth]{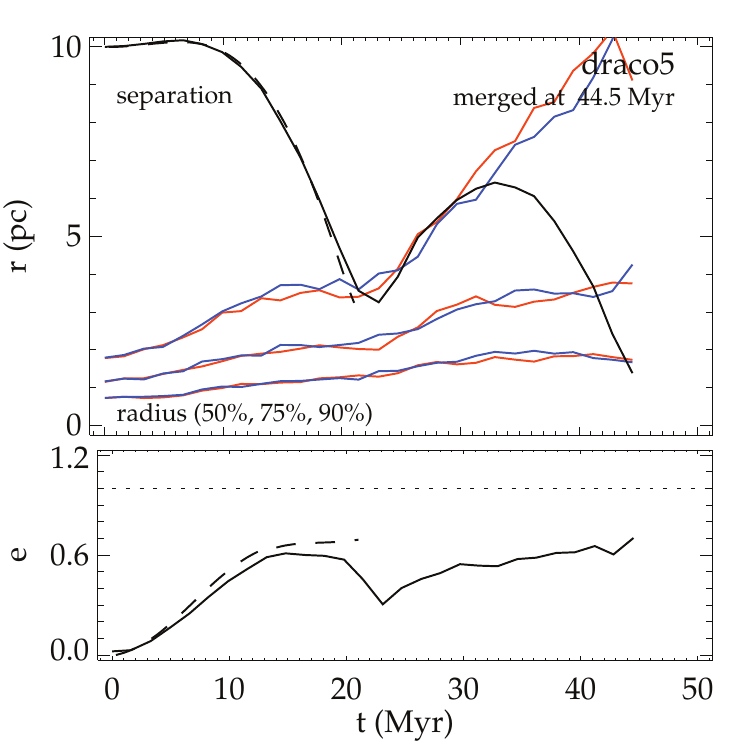}
\includegraphics[width=0.245\textwidth]{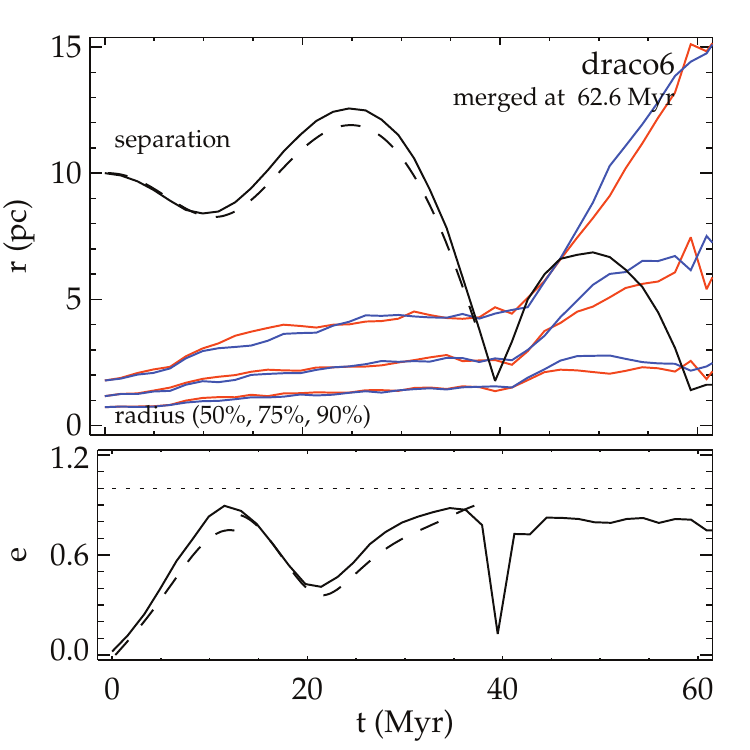}
\includegraphics[width=0.245\textwidth]{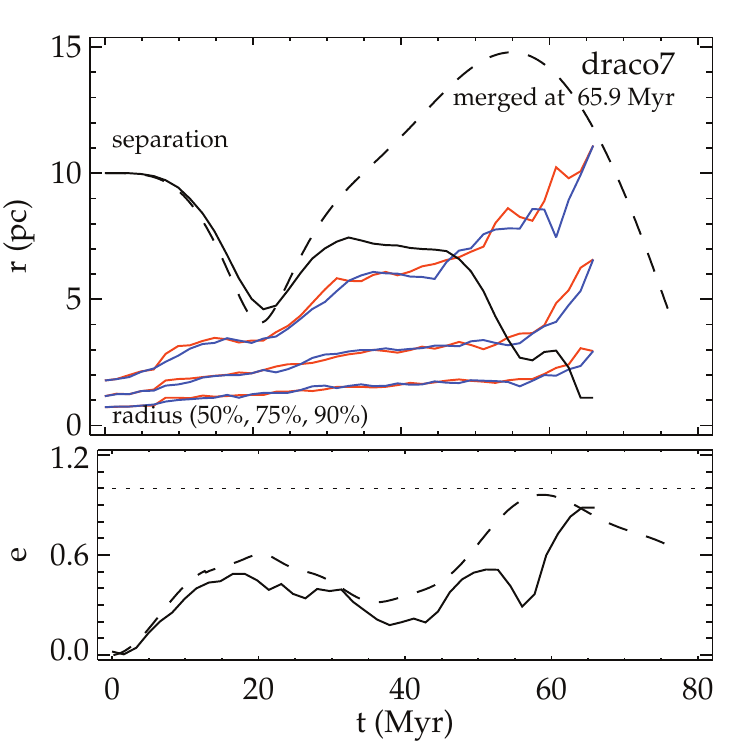}
\includegraphics[width=0.245\textwidth]{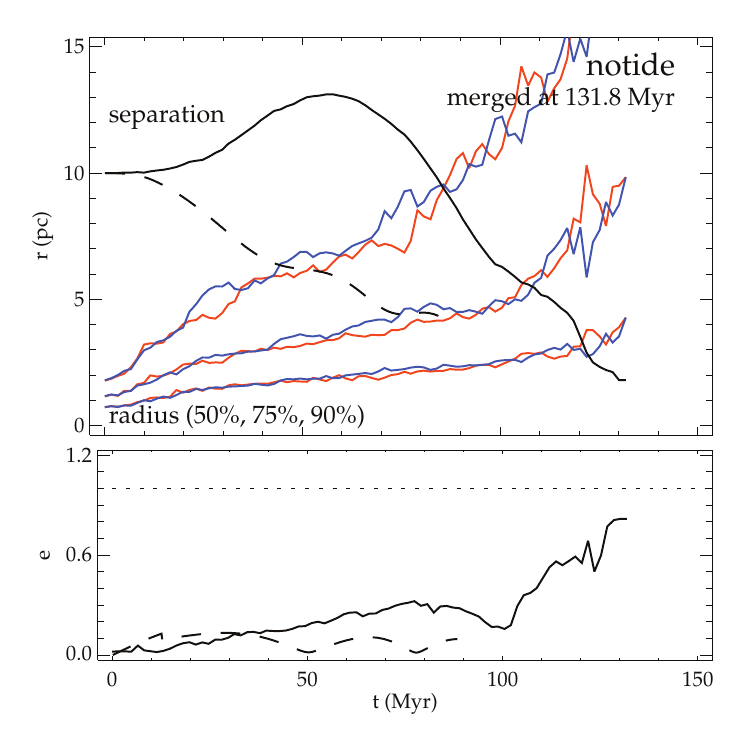}
\caption{Orbital evolution of various models, represented by their separations ($r$) and eccentricities ($e$). Results from $N$-body simulations (solid lines) are compared to semi-analytic results (dashed) with identical initial conditions. The red and blue curves indicate the 50\% (lower curves), 75\% (middle curves) and 90\% (upper curves) Lagrangian radii of the two star clusters,  and provide insight into the effect of tidal interactions between the two clusters.}
\label{radius}
\end{figure*}

Although the semi-analytic models described above provide fast and valuable insights about the evolution of binary star clusters, they are unable to model the merging process and lack the details that are required for a deeper investigation of the dynamics. $N$-body simulations, although slower, can give us the necessary information about the merger process, cluster expansion, cluster rotation, and about the escaping stellar population. The results of the $N$-body simulations of the eight binary star clusters listed in \autoref{initial} are shown in \autoref{position} and \autoref{radius}, while the time scales at which the mergers take place are listed in \autoref{merger}.

The isolated binary cluster (\notide) merges into one cluster, as the binary always loses its angular momentum  due to escaping stars, which are ejected from both clusters as a result of internal dynamical mechanisms  \citep{sugimoto1989}. Mass loss due to stellar evolution at early times also widens the binary cluster separation and postpones orbital decay \citep{spz2007}. The two clusters in model \notide{} make over than two revolutions before merging after more than 130~Myr.

All but one of the \draco{} binary cluster models lead to a merger, as predicted by our semi-analytic models (see \autoref{semi}). The only exception is model \draco3, which has a planar prograde orbit with initial conditions ($\Omega,\theta$) = ($0^\circ, 90^\circ$). Note that, since $i=0^\circ$, the latter is equivalent to ($\Omega,\theta$) = ($90^\circ, 0^\circ$). \draco3 suffers from a strong outward tidal force within half an orbital period. The \draco1 system is identical to \draco3, apart from the initial phase angle, and merges after 92.2~Myr. This demonstrates that the initial phase angle is an important contributing factor to the dynamical fate of a binary star cluster.

The \draco1 model provides a unique case where the initial orbital condition lies near the boundary of the merging and separation region (see \autoref{semi}), but finally merges at age $t\approx92$~Myr. Initially, this binary has a prograde orbit (counter-clockwise) with one component located at apogalacticon. The external tidal field stretches the distance between two clusters, but to such an extent that it almost disrupts the binary. A maximum separation of $r\approx20$~pc is achieved at $t\approx25$~Myr, almost half of the initial orbital period. The binary changes its orbital spin direction and subsequently experiences a nearly head-on collision that ends in a merger. At the start of the collision, the relative velocity between two clusters is about 1.5~km\,s$^{-1}$, slightly above the velocity dispersion of the member stars of each of the two clusters. However, the tidal interaction between the two clusters does not severely change their structure as shown in \autoref{radius}, where an abrupt increase of clusters' radii is not observed.

Models with initially retrograde orbits (\draco2 and \draco4) have relatively short merging time scales as the tidal force accelerates the merging process. \draco2, which starts at $\theta=0\drj$, merges faster than \draco4. This is in contrast to the simulation results form \cite{innanen1972}, who find that binary stellar systems are more likely to survive when they have retrograde orbits with respect to the systematic orbits around the galaxy. 
Contrary to the results of the {\notide} model, both \draco2 and \draco4 obtain a high eccentricity relatively early on, which is caused by the external tidal force. These eccentric orbits both lead to pericentre passage at $t\approx20$~Myr, but with different minimum distances. As the tidal force has a  stronger contribution to the evolution of \draco2, the minimum distance between the components becomes as small as 3~pc. At this point, tidal circularisation (decrease in eccentricity) and tidal distortion (increase of the cluster halo radii) are observed. However, the half-mass radii ($r_{50\%}$) are only marginally affected by the close encounter.

The orbital evolution of the perpendicular models (\draco$5-7$) presents another effect of the tidal influence on binary star clusters. Both \draco5 and \draco7 start at the Galactic plane while the two clusters in \draco6 are initially located above and below Galactic plane. In all these cases, the tidal force tumbles the orbital axis of all these binary star clusters, up to the moment of the merger events. As indicated by the angular momentum components of the merger remnants (see \autoref{merger}), there is no preferred direction to which the orbital spins are directed by the tide. Each component of the tidal force acts independently according to the instantaneous position and velocity of the subject body. Furthermore, oscillating orbits about Galactic plane are observed in all three models. The vertical component of tidal force is responsible for the clusters' separations to bounce up after a close encounter. Consequently, binary clusters with initially perpendicular orbits have moderately longer lifetimes, before experiencing a merger.
An abrupt expansion of the cluster radius occurs just after a close encounter. Even very close encounters (or collisions) with an impact parameter below 2~pc occur during the evolution of \draco6, causing a steep increase in the clusters' radii. As predicted by the impulse approximation, an encounter with smaller impact parameter yields a larger energy change, which in turn expands the cluster radius at a higher rate.


\subsection{Comparison between the semi-analytic models and $N$-body models}

Our semi-analytic model accounts for mass loss, angular momentum loss and the external tides of a galaxy, while the more realistic $N$-body simulations account for all dynamical processes. A comparison between these two approaches is necessary to characterise the validity of the (much faster) semi-analytic model. Such a comparison, in terms of separation and eccentricity, is presented in \autoref{radius}.

In most cases, the results of the semi-analytic models are in very good agreement with the orbital evolution obtained using the $N$-body simulations, at least in the time span before a close encounter occurs. Minor differences are generally present, in which the semi-analytic approach underestimates the separations and overestimates the eccentricities. These differences may be due to the variation of the coefficient $l$ in \autoref{angmo} as the orbital separations become so large and the clusters graze the tidal radius that is determined by the Galactic environment. The value of this coefficient also depends on the orbital eccentricity \citep{nariai1976}.
Larger discrepancies appear for the cases of \draco7 and \notide. In \draco7, there is a close encounter that results in a more eccentric orbit. Since the semi-analytic model does not account the tidal interaction between the two clusters (synchronisation and dynamical friction), it gives a less accurate prediction of the merger timescale. In \notide, the semi-analytic model predicts an immediate shrinkage of the orbital separation, while the $N$-body model experiences expansion. The difference arises from the evaporation rate of the isolated model, which cannot be represented by \autoref{disrupt}, even with different values of $t_0$.


\subsection{Rotating merger remnants}

\begin{figure}
\centering
\includegraphics[width=0.42\textwidth]{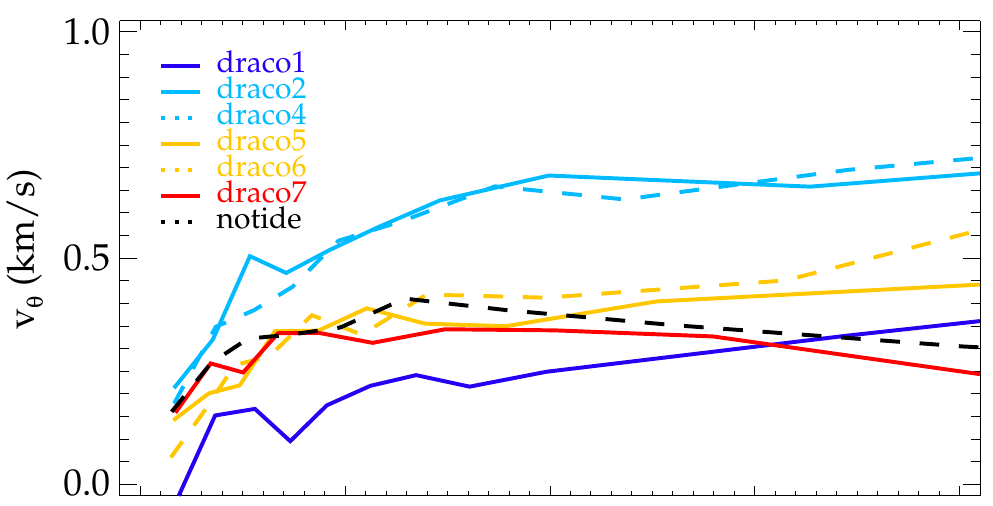}\\
\includegraphics[width=0.42\textwidth]{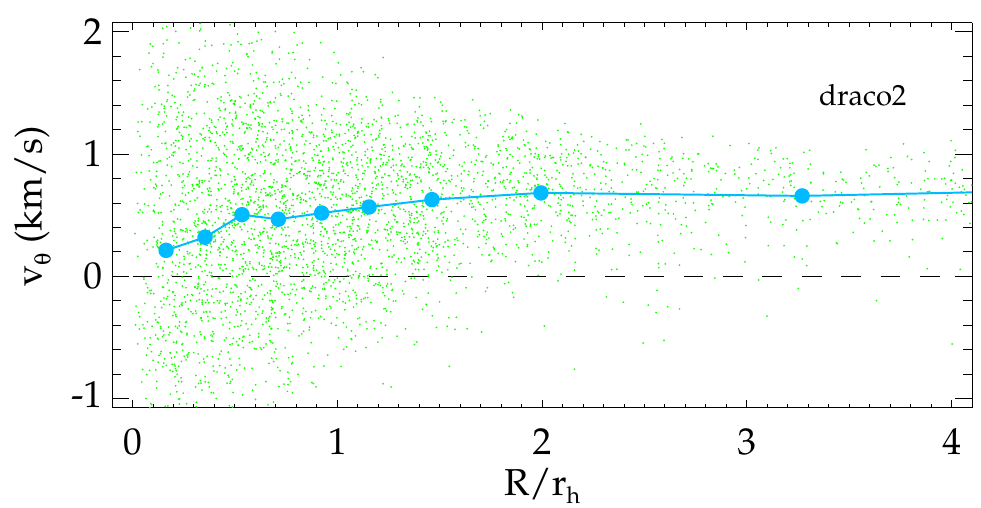}
\caption{{\em Top}: rotation curves of the different $N$-body models, 10~Myr after the merger has taken place (cf. \autoref{initial} and \autoref{merger}). The horizontal and vertical axes represent the cylindrical radius (\autoref{cylindricalradius}) and rotational velocity (\autoref{rotationalvelocity}), respectively. {\em Bottom}: rotational velocities of the bound stars in the \draco2 model 10~Myr after the merger has taken place. The corresponding rotation curve is indicated with the solid curve. }
\label{rcurve}
\end{figure}

As discussed in previous works \citep[e.g.][]{sugimoto1989,marcos2010}, a merger of a binary star cluster transforms orbital angular momentum into rotational angular momentum, and thus produces a rotating merger remnant. We therefore expect that the rotational nature of the remnants strongly depends on a combination of the orientation of the initial binary orbit and and the external tidal field.
To understand the rotational nature of the merger remnants, we first define the net angular momentum of a star cluster, which is the sum of the angular momenta of the member stars:
\begin{equation}
\vec{L}=\sum_i{m_i\vec{r}_i\times \vec{v}_i}  \  .
\end{equation}
Here, the positions $\vec{r}_i$ and velocities $\vec{v}_i$ are relative to the centre of mass of the cluster, and stellar members are identified as those that are located at a distance less than two tidal radii from the cluster centre of mass \citep[the default criterion in \nbody; see][]{aarseth2003}. If the cluster has significant rotation (velocity anisotropy), then the net angular momentum $L=|\vec{L}|$ is non-zero, and the direction of the angular moment direction $\hat{L}\equiv\vec{L}/|L|$ defines the rotation axis of the cluster. The values of the three components of $L=(L_x,L_y,L_z)$ for the different models are listed in \autoref{merger}. 
Subsequently, we calculate the cylindrical radius $\mathcal{R}$ (the distance of the star to the rotational axis:
\begin{equation}
\mathcal{R}_i=\dfrac{|\vec{L}\times\vec{r}_i|}{|\vec{L}|} \ .
\label{cylindricalradius}
\end{equation}
We also calculate the rotational velocity $v_{\theta}$ of each star, which we define as the velocity component perpendicular to rotational axis, projected onto the equatorial plane (i.e., the plane with normal vector $\hat{L}$). This quantity is calculated as follows
\begin{equation}
v_{\theta,i}=\dfrac{\left(\vec{r}_i\times\vec{v}_i\right)\cdot\hat{L}}{\mathcal{R}_i}.
\label{rotationalvelocity}
\end{equation}
The rotational velocities of the stars in the merger remnants are shown in \autoref{rcurve}. Rotation can also be represented in the form of a rotation curve that consists of the average value of rotational velocity as a function of the cylindrical radius. 
Clusters with significant rotation show an asymmetric distribution of $v_{\theta}$ about zero. The dispersion of $v_{\theta}$ is large near the centre while the rotation curve almost resembles rigid rotation up to the half-mass radius \citep[see also][]{kim2002}. Outside that radius, the average value of $v_{\theta}$ is expected to drop radially \citep[e.g.][]{kim2004, kim2008, hong2013}. However, some merger products simulated in this study do not reveal a clear decrease in the rotational velocity at large cylindrical radii, as expected from single rotating star clusters. Dynamical processes experienced by the binary clusters prior to the merger might be the main cause of this difference.

The degree of rotation does not scale with the time at which the merger occurs, as each of the models went through different trajectories before the merger. \draco1 has the smallest degree of rotation as it is the result of a head-on collision. The increased rotational velocity of stars in the cluster halo stars  is mainly a result of the tidal field. On the other hand, the spiral-in orbital decay of both \draco2 and \draco4 induce the fastest rotation in the merger remnant.

Perpendicular models produce merger remnants with a moderate amount of rotation. \draco7 exhibits a slightly different rotation curve as compared to the other two perpendicular models (\draco5 and \draco6). Instead of an increase with radius, the rotation curve conforms to a more Keplerian trend, similar to that of the isolated model \notide{} \citep[see also][]{makino1991}. This is a result of the peculiar orientation of \draco7's rotational axis with respect to the Galactic plane.

The amplitudes of the velocity curves of merger remnants decrease over time, as angular momentum is removed from escapers through evaporation and ejection of member stars, which is a well-studied mechanism of angular momentum removal from rotating star clusters \citep[see, e.g.,][and references therein]{akiyama1989, spurzem1991, einsel1999, ernst2007, hong2013}. However, a systematic non-zero rotational velocity of halo stars is preserved over long periods of time, due to the presence of the external tidal force.


\subsection{Evaporation rates}

Merging binary clusters are subject to a higher rate of mass loss due to a combination of several mechanisms. First, close encounters and collisions between two clusters increase the internal energy of each cluster, which in turn induces radial expansion and increases the mass loss rate. Numerical simulations of unequal-mass binary clusters in isolation conducted by \cite{oliviera2000} demonstrated that colliding binary star clusters eject a fraction ($\leq 3\%$) of their total mass into the field. This fraction is small, but the external tidal field in our simulations substantially increases the fraction of escaping stars. During and shortly after the merger process, the merging \draco{} models  lose $\sim 10\%$ of their stars in an outburst.

The second mechanism is the enhanced mass loss due to cluster rotation itself. \cite{ernst2007} performed $N$-body simulations of star clusters with different rotational speeds, considering both prograde and retrograde rotation relative to the orbit of the star cluster around the host galaxy. They found a significant difference in the mass loss rates of rotating and non-rotating star clusters. Moreover, star clusters with prograde rotation suffer from higher mass loss rates as compared to retrograde ones. The reason behind this difference is that many stars with retrograde orbits are subject to a 'third integral' of motion, which hinders their escape \citep{fukushige2000}.

\begin{figure}
\centering
\includegraphics[width=0.47\textwidth]{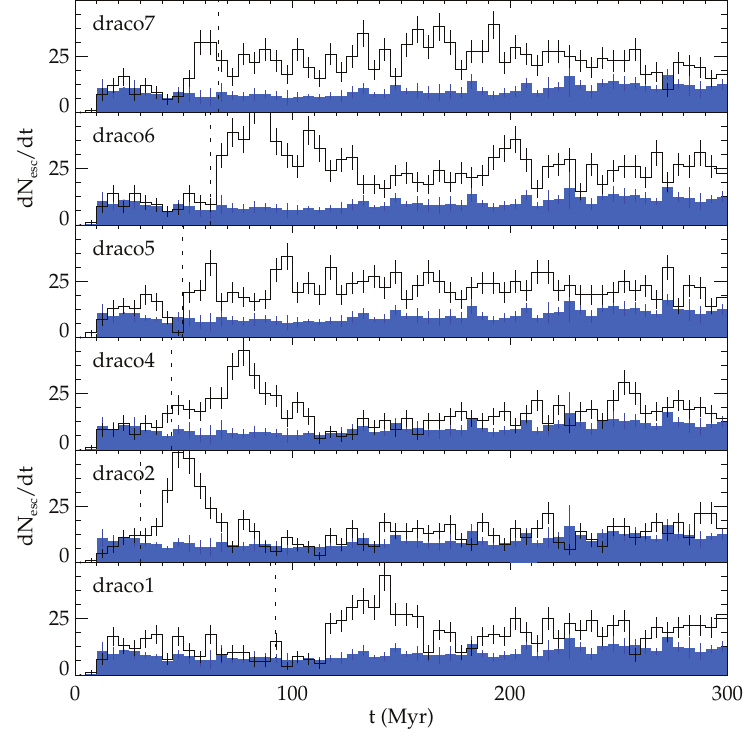}
\caption{Evaporation rates of the merger remnants of the \draco{} binary clusters (open histograms), in units of number of escapers per time interval of 5~Myr. The dashed vertical lines mark the time of merger for each binary cluster. Poissonian error bars are indicated for the escape rates of the \draco{} binary star clusters. For comparison we also show the evaporation rates of the \textsc{single} cluster model (solid histograms), where the error bars represent the standard deviation of five runs.}
\label{evaporation}
\end{figure}

Clear evidence of these two mechanisms can be found in our simulations; see \autoref{evaporation}. For self-similar models, such as non-rotating single star clusters, the evaporation rate is more or less constant over time \citep[e.g.][]{lee1987, gieles2011, wang2015enc}. On the other hand, a significant increase in the evaporation rate occurs during and after a close encounter between two star clusters, both for encounters that lead to a merger, as well as for encounters that do not. An abrupt increase in the evaporation rate (or outburst) occurs $\sim20$ Myr after a close encounter. The latter timescale is mainly a result of our modelling: potential escapers with a kinetic energy above the local escape velocity require some time to reach the adopted escape radius ($r=2r_J$), before being considered as escaping stars. Note that there is only one close encounter in \draco1 that leads to a merger. The other models experience multiple close encounters before their final merger takes place. 

A further inspection of \draco1, \draco2 and \draco4 reveals outburst profiles which show that a close encounter or merger results in the ejection of  $\sim 15\%$ of the member stars. After the encounter, the system find its equilibrium back, and the evaporation rate drops to a constant value. A merger of a binary cluster with an initially planar orbit produces rotating clusters with a retrograde spin orientation (with respect to their Galactic orbits), indicated by $L_z<0$ (\autoref{merger}). Therefore, the merger remnants have a somewhat higher evaporation rates as compared to those of the non-rotating models. 
Models with initially perpendicular orbits exhibit different trends. The evaporation rate increase during the close encounter phase, but this rate remains at least a factor two of the rate of the single, non-rotating cluster, which reduces the lifetime of the cluster. This higher evaporation rate is related to the rotation of the merger remnant, whose axis is tilted with respect the normal vector of the Galactic plane. The vertical component of the tide reduces the angular momentum in the $x$ and $y$ directions, leaving only a non-zero $L_z$. During the entire process, a large fraction of the orbital angular momentum needs to be dissipated, and this occurs mainly through the evaporation of member stars.


\section{Discussion} \label{discussion}

Binary star clusters can form through several mechanisms, including primordial fragmentation or fission \citep{elmegreen1996,fujimoto1997,theis2002,bekki2004} and sequential star formation \citep{brown1995,bik2010}. The two clusters may also form independently and pair up into a binary through tidal capture \citep{vandenbergh1996}, or through interactions with other clusters in dense regions, similar to binary formation through capture of stellar-mass objects in clusters \citep[e.g.,][and references therein]{kouwenhoven2010,perets2012}. Fission of giant molecular clouds triggered by another passing perturber is considered as the most probable way to produce binary clusters. \cite{dieball2002} found that cluster pairs in the LMC and SMC tend to be coeval and have mass ratios near unity, and such characteristics are consistent with a fission mechanism of binaries \citep[see, e.g.,][]{kouwenhoven2009}. 

Binary clusters in the Milky Way galaxy may have formed through a similar process, where globular clusters passing through the disk may be able to perturb giant molecular clouds (GMCs), triggering the formation of open clusters \citep{marcos2014}.
The resulting population includes binary star clusters with bound (or unbound) orbits that can be oriented in any direction, depending on the motion of perturbing body. There is no preferential orbital orientation with respect to the Galactic plane. However, it is hard to determine the orbital orientation of any cluster pair observationally, since the kinematic data (proper motions and radial velocities) are very difficult to obtain. On the other hand, a rough estimate of the minimum orbital inclination, $i_{\text{min}}$, can be calculated using the position and distance of the cluster sample as $\tan i_{\text{min}}=z/r$, where $r$ expresses the binary separation, and $z$ is the separation component perpendicular to the Galactic plane. This parameter can be used as a proxy to understand the behaviour of binary clusters in Galactic tidal field.

\cite{marcos2010} identified seven pairs of star clusters as bound systems, according to the analysis of their tidal radii and separations. These are the pairs NGC\,1976-1981, ASCC,20-ASCC\,16, NGC\,3324-NGC\,3293, Collinder\,197-ASCC\,50, NGC\,6250-Lynga\,14, NGC\,6613-NGC\,6618, and Trumpler\,22-NGC\,5617. These pairs have minimal inclination between $15^{\circ}$ and $45^{\circ}$, as computed using positions and distances listed in \emph{The Open Cluster Database} \citep[WEBDA; see][]{mermilliod2003}. Among those pairs, Trumpler\,22-NGC\,5617 is the oldest cluster pair with an estimated age of 82~Myr, while the minimal orbital inclination is about $37^{\circ}$, which is rather high. As demonstrated with our simulations, the lifetime of binary clusters depends on the orbital orientation, but we can expect binary clusters with high orbital inclinations to have a longer lifetime before merging or separating.


\section{Conclusions}
\label{conclusion}

We present a dynamical study on the dynamical evolution of binary star clusters with different initial orbital orientations with respect to the Galactic plane. We have carried out numerical simulations to study the internal and external processes involved in the processes that lead to the mergers or separations  of binary star clusters. Our main conclusions are summarised as follows:

\begin{enumerate}

\item We have developed a fast semi-analytic model to study the evolution of arbitrarily oriented binary star clusters. The model accounts for cluster mass loss, angular momentum loss due to escaping stars, and a galactic tidal field. The model does not include the mutual tidal interaction between the two star clusters, which is the most important process prior to an imminent merger. A comparison with $N$-body simulations indicates that this approach works good, as long as Roche lobes of the clusters do not overlap. The merger process and subsequent evolution are studied using the direct $N$-body package \nbody.

\item Binary star clusters with identical initial semi-major axes and eccentricities (i.e., identical orbital energy and angular momentum), may experience completely different dynamical fates, depending on their initial orientations relative to the Galaxy, and their initial orbital phase. Within the strong influence of tidal field, the orbital evolution of binary clusters may involve orbital reversal, spiral-in and vertical oscillation about Galactic plane, prior to obtaining their final configurations (merger or separation)

\item Before a merger occurs, the components of a binary cluster system typically experience at least one close encounter that increases clusters' internal energy and expands their radii. At this point, tidal friction becomes an increasingly important process that may ultimately result in a merger.

\item Mergers of binary star clusters generally result in rotating clusters that inherit a fraction of the orbital angular momentum. The rotation axis of this merger remnant is determined by the orbital orientation prior to the merger, and is typically retrograde with respect to the cluster's orbit around the galaxy. The latter may differ substantially from the initial orbital orientation, due to interaction with the external tidal field.

\item Within a tidal field, a merger between two star clusters is typically associated with an outburst that ejects $\sim10\%$ of the member stars into the field. This value is significantly higher than what is found for isolated mergers \citep{oliviera2000}. Merger remnants with retrograde rotation exhibit a nearly constant evaporation rate, which is slightly higher than that of a single, non-rotating cluster. Merger remnants of binary clusters with an initially highly tilted orbit also tend to have a retrograde spin orientation, but with the rotation axis more or less perpendicular to the Galactic plane. Consequently, these kinds of systems suffer from higher evaporation rates: almost twice the rate of a single cluster with a similar mass and number of stars. 

\end{enumerate}
Our results underline several important aspects of binary clusters that affect the evolution of a star cluster population in the Milky Way and other galaxies. The results of our study can in principle be tested observationally, allowing us to further constrain the scenario of binary cluster formation through the fission mechanism, and the predicted relationship between binary star cluster lifetime and the orbital orientation with respect to the disk plane of the host galaxy.

The main purpose of our study was to characterise the strong relation between the initial orbital orientation and initial phase angle of binary star clusters evolving in an external tidal field. Although our semi-analytic model makes several assumptions, it can be improved to match the initial conditions of observed binary star clusters. Moreover, recent developments in $N$-body tools now make it possible to study much more massive binary cluster systems (including globular clusters) directly, using \textsc{nbody6++gpu} \citep{wang2015} with arbitrary  time resolution \citep{cai2015}, or using the multi-scale, multi-physics \textsc{amuse} package \citep[e.g.,][]{pelupessy2013,spz2013}.


\section*{Acknowledgments}

We wish to express our gratitude to the anonymous referee, for providing constructive comments that helped to improve this paper.
This work was carried out during the research visit of R.P. to the Kavli Institute for Astronomy and Astrophysics at Peking University, which was partially supported by Leids-Kerkhoven-Bosscha Fonds (LKBF) and the Ministry of Education of Indonesia through Biaya Operasional Perguruan Tinggi (BOPTN). 
R.P., M.I.A., and H.R.T.W. were supported by the Ministry of Research and Higher Education through Hibah Riset Desentralisasi Research grant (310b/I1/C01/PL/2015).
M.B.N.K. was supported by the Peter and Patricia Gruber Foundation through the PPGF fellowship, by the Peking University One Hundred Talent Fund (985), and by the National Natural Science Foundation of China (grants 11010237, 11050110414, 11173004 and 11573004). This publication was made possible through the support of a grant from the John Templeton Foundation and National Astronomical Observatories of Chinese Academy of Sciences. The opinions expressed in this publication are those of the author(s) do not necessarily reflect the views of the John Templeton Foundation or National Astronomical Observatories of Chinese Academy of Sciences. The funds from John Templeton Foundation were awarded in a grant to The University of Chicago which also managed the program in conjunction with National Astronomical Observatories, Chinese Academy of Sciences.


\bibliographystyle{mnras}
\bibliography{paper}


\appendix

\section{Initial conditions for the binary star clusters}
\label{appendix1}

\begin{figure}[h]
\centering
\includegraphics[width=0.45\textwidth]{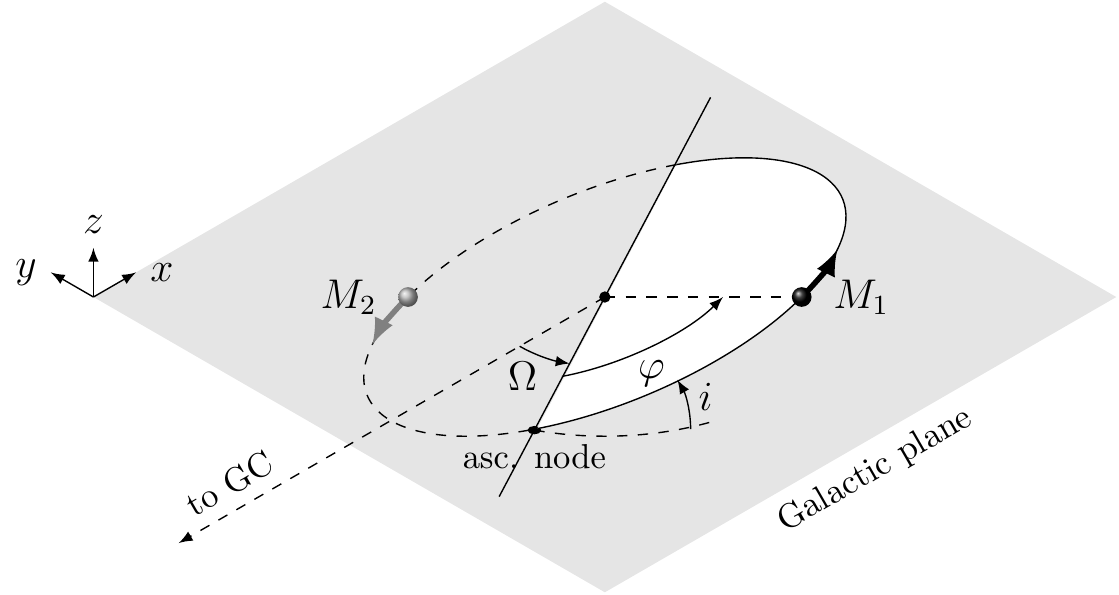}
\caption{Illustration of the orbital orientation of the binary star cluster, relative to the Galactic centre (GC).}
\label{orbital}
\end{figure}

We construct a system of two star clusters with masses $M_1$ and $M_2$, respectively. The clusters form a binary system with semi-major axis $a$ and eccentricity $e$. The longitude of the ascending node $\Omega$, the inclination angle $i$ and the argument of pericentre $\phi$ define the orbital orientation in space, with respect to the host galaxy. Finally, the phase angle (or true anomaly), $\theta$, defines the initial positions of the clusters in the orbit (see \autoref{orbital}).
To determine the clusters' positions and velocities at $t=0$, their relative distance $r_{\text{rel}}$ and velocity $v_{\text{rel}}$ are computed:
\begin{equation}
r_{\text{rel}}=\dfrac{a(1-e^2)}{1+e\cos{\theta}}
\quad \quad
v_{\text{rel}}=\left[GM\left(\dfrac{2}{r_{\text{rel}}}-\dfrac{1}{a}\right)\right]^{1/2} \ ,
\end{equation}
where $M=M_1+M_2$ is the total mass of the binary system and $G$ is the gravitational constant.
It is convenient to adopt a coordinate system in which both star clusters are aligned along $x$-axis, and the centre of mass is located at the origin. One cluster moves in the positive $y$-direction, while the other cluster moves towards negative $y$, resulting in a prograde or retrograde orbit, depending on the orbital orientation. The angle $\alpha$ between the position vector and the velocity vector depends on both the orbital eccentricity and the phase angle. Using the conservation of angular momentum $L$, it is straightforward to obtain the following relation:
\begin{equation}
\alpha=\sin^{-1}\left[\dfrac{a^2(1-e^2)}{2ar_{\text{rel}}-r_{\text{rel}}^2}\right]^{1/2}.
\end{equation}
Thus, the position and velocity of each cluster become
\begin{align}
&\vec{r}_1=\left(1,0,0\right)qr_{\text{rel}}
&\vec{v}_1=\left(\sin{\alpha},-\cos{\alpha},0\right)qv_{\text{rel}} \\
&\vec{r}_2=\left(-1,0,0\right)Qr_{\text{rel}}
&\vec{v}_2=\left(-\sin{\alpha},\cos{\alpha},0\right)Qv_{\text{rel}}
\end{align}
where $q \equiv M_2/M$ and $Q \equiv M_1/M$. These positions and velocities can then be transformed into the desired orbital orientation:
\begin{equation}
\vec{r}'=\vec{r}
\left[
\begin{array}{p{8mm}p{6mm}p{0.5mm}}
$\cos\Omega$ & $\sin\Omega$ & 0\\
$-\sin\Omega$ & $\cos\Omega$ & 0\\
0 & 0 & 1
\end{array}
\right]
\left[
\begin{array}{p{0mm}p{7mm}p{6.4mm}}
1 & 0 & 0\\
0 & $\cos{i}$ & $\sin{i}$ \\
0 & $-\sin{i}$ & $\cos{i}$
\end{array}
\right]
\left[
\begin{array}{p{7mm}p{5mm}p{1mm}}
$\cos\varphi$ & $\sin\varphi$ & 0\\
$-\sin\varphi$ & $\cos\varphi$ & 0\\
0 & 0 & 1
\end{array}
\right] \ ,
\end{equation}
where $\varphi\equiv\theta+\omega$. A similar transformation can also be applied to the velocity vector.

\bsp

\label{lastpage}

\end{document}